\documentclass[10pt,preprint2]{aastex}
\usepackage{amsmath}
\usepackage{natbib}
\usepackage{float}
\usepackage{multirow}
\usepackage{pdflscape}
\usepackage[hypertex]{hyperref}
\usepackage[margin=2.5cm]{geometry}

\begin{document}

\title{Measuring the Galaxy Cluster Bulk Flow from WMAP data}
\author{S.J. Osborne \altaffilmark{1} \altaffilmark{2}, D.S.Y. Mak \altaffilmark{3}, S.E. Church \altaffilmark{1} \altaffilmark{2}, E. Pierpaoli \altaffilmark{3}}
\altaffiltext{1}{Stanford University, 382 Via Pueblo, Varian Building, Stanford, CA, 94305-4060}
\altaffiltext{2}{Kavli Institute for Particle Astrophysics and Cosmology at Stanford University, 452 Lomita Mall, Stanford, CA, 94305-4085}
\altaffiltext{3}{University of Southern California, Los Angeles, CA, 90089-0484}

\begin{abstract}
We have looked for bulk motions of galaxy clusters in the WMAP~7 year data. We isolate the kinetic Sunyaev-Zeldovich (SZ) signal by filtering the WMAP Q, V and W band maps with multi-frequency matched filters, that utilize the spatial properties of the kinetic SZ signal to optimize detection. We try two filters: a filter that has no spectral dependence, and a filter that utilizes the spectral properties of the kinetic and thermal SZ signals to remove the thermal SZ bias. We measure the monopole and dipole spherical harmonic coefficients of the kinetic SZ signal, as well as the $\ell=2-5$ modes, at the locations of 736 ROSAT observed galaxy clusters. We find no significant power in the kinetic SZ signal at these multipoles with either filter, consistent with the $\Lambda$CDM prediction. Our limits are a factor of $\sim \! 3$ more sensitive than the claimed bulk flow detection of~\citet{2009ApJ...691.1479K}. Using simulations we estimate that in maps filtered by our matched filter with no spectral dependence there is a thermal SZ dipole that would be mistakenly measured as a bulk motion of $\sim \! 2000-4000$ km/s. For the WMAP data the signal to noise ratio obtained with the unbiased filter is almost an order of magnitude lower.
\end{abstract}


\section{Introduction}

Verifying the depth of convergence of the $627$ km/s motion of the Local Group (LG) with respect to the frame defined by the Cosmic Microwave Background (CMB) has been a long-standing goal of observational cosmology. Cosmographers seek to identify the region in which the local universe is at rest with respect to the CMB frame, and the LG motion has converged to the CMB dipole direction. This region would contain all of the mass sources responsible for the motion of the LG and would define a sample of the universe representative of the whole. The coherent motion of matter caused by gravitational potentials is termed a bulk flow~\citep[e.g.,][]{1995PhR...261..271S}; if the flow covers the entire universe it is sometimes referred to as a dark flow~\citep[e.g.,][]{2008ApJ...686L..49K}. Such a flow is equivalent to an intrinsic dipole of the universe. Models of inflation can be constructed to explain such a measurement~\citep[e.g,][]{1991PhRvD..44.3737T,1994PhRvL..73.1582K}.

To date there is no consensus on the depth of convergence. The CMB dipole magnitude and direction have been precisely measured by WMAP: ($3.355 \pm 0.008$) mK in the direction of galactic longitude $l=263.99^{\circ} \pm 0.14^{\circ}$ and latitude $b=48.26^{\circ} \pm 0.03$~\citep{2010arXiv1001.4744J}. The LG motion towards the Great Attractor was measured by~\citet{1987ApJ...313L..37D} in the direction of $l=312^{\circ}$, $b=6^{\circ}$ and found to be coherent out to $30 h^{-1}$ Mpc and near zero by $40 h^{-1}$ Mpc. Using a full-sky peculiar velocity survey, with a depth ranging from $80 h^{-1}$ Mpc to $110 h^{-1}$ Mpc,~\citet{1994ApJ...425..418L} found a dipole toward ($l=220^{\circ}$, $b=-28^{\circ}$) $\pm 27^{\circ}$, inconsistent with the CMB dipole direction at $99.99\%$ confidence and generated by mass concentrations beyond $100 h^{-1}$ Mpc. An analysis of IRAS indicated a dipole direction within $13^{\circ}$ of the CMB dipole that has not fallen to zero by $300 h^{-1}$ Mpc~\citep{2000MNRAS.314..375R}.~\citet{2004ApJ...608..721K} used an all-sky X-ray selected cluster catalog and found that the direction of the LG's peculiar velocity is well aligned with the CMB as early as the Great Attractor region $40 h^{-1}$ Mpc away. However, most of the~\citet{2004ApJ...608..721K} dipole signal is attributed to the Shapley supercluster $150 h^{-1}$ Mpc away, as well as a handful of massive clusters behind our galaxy.

Using $56$ SMAC clusters within $120 h^{-1}$ Mpc,~\citet{2004MNRAS.352...61H} found a bulk flow of $687 \pm 203$ km/s toward $l=260^{\circ} \pm 13^{\circ}$, $b=0^{\circ} \pm 11^{\circ}$ that did not drop off with depth. They found that it could not be caused by the Great Attractor, but could be caused by the Shapley Concentration (at marginal significance). They argue that multiple data sets exclude convergence to the CMB frame by $60 h^{-1}$ Mpc, and that at depths of $60-120 h^{-1}$ Mpc the flow is limited to 600 km/s. Using a new method of optimally weighting peculiar velocities,~\citet{2009MNRAS.392..743W} compiled all major peculiar velocity surveys (including SMAC) and found them to be highly consistent. Within a region of radius $\sim \! 100 h^{-1}$ Mpc they find a flow of $407 \pm 81$ km/s toward $l=287^{\circ} \pm 9^{\circ}$, $b=8^{\circ} \pm 6^{\circ}$ implying that $50\%$ of the LG motion is generated beyond this depth. Extending this analysis \citet{2010MNRAS.407.2328F} find a consistent flow, $416 \pm 78$ km/s toward $l=282^{\circ} \pm 11^{\circ}$, $b=6^{\circ} \pm 6^{\circ}$. These results are found to be in disagreement with the $\Lambda$CDM model with WMAP~5 year cosmological parameters at a high confidence level.

Using the 2MASS Redshift Survey~\citet{2010ApJ...709..483L} find that less than half of the amplitude of the CMB dipole is generated within $\sim \! 40 h^{-1}$ Mpc, and that most of the amplitude of the dipole is recovered by $120 h^{-1}$ Mpc, although the directions of the two flows do not agree.~\citet{2010arXiv1011.6292C} use Type Ia supernovae (SNe 1a) to find that there is a bulk flow of around 260 km/s at $\sim \! 180 h^{-1}$ Mpc, which disagrees with $\Lambda$CDM at the $1-2 \sigma$ level. However, at $\sim \! 435 h^{-1}$ Mpc they find improved agreement between the SNe Ia data and the isotropic $\Lambda$CDM model.

The kinetic Sunyaev-Zeldovich (SZ) effect was first used to place a limit on the bulk flow velocity by~\citet{2003ApJ...592..674B} using $10$ clusters between $300-600 h^{-1}$ Mpc. They found no detection but limited the flow in the direction of the CMB dipole to $\le 1410$ km/s at $95\%$ confidence. \citet{2000ApJ...536L..67K} found that by utilizing the kinetic SZ effect in WMAP data, flows as small as 200 km/s (or 30 km/s for the Planck experiment~\citep{cite:bluebook}) could in principle be measured. Using the WMAP~5 year data~\citet{2008ApJ...686L..49K}~\citep[also][hereafter KAKE]{2009ApJ...691.1479K} found a coherent dipole out to at least a distance of $300 h^{-1}$ Mpc. When all clusters in their sample beyond $300 h^{-1}$ Mpc are combined the dipole aligns well with the CMB dipole and is within $6^{\circ}$ of the dipole found by~\citet{2009MNRAS.392..743W}. However, the magnitude of the KAKE dipole is considerably larger, estimated to be between $600-1000$ km/s and is detected on much larger scales with most of the signal coming from a region between $120-600 h^{-1}$ Mpc away. Using a larger cluster sample~\citet{2010ApJ...712L..81K} find that the flow has constant velocity out to approximately $575 h^{-1}$ Mpc. However, using the same method as KAKE,~\citet{2009ApJ...707L..42K} do not find a significant detection of a bulk flow. They find that residual CMB in the filtered KAKE maps, that is correlated between the WMAP channels and not accounted for in the errors, decreases the significance of the cluster dipole.~\citet{2010ApJ...719...77A} present an analysis of the errors in the cluster dipole measurement. They do not reproduce the errors in~\citet{2009ApJ...707L..42K} unless the monopole and dipole are removed from the full sky CMB map instead of only in the region outside of the galactic mask, suggesting this as a reason for the null result of~\citet{2009ApJ...707L..42K}.

We measure the galaxy cluster peculiar velocity distribution using the WMAP~7 year data. The main difference between our method and that of KAKE and~\citet{2009ApJ...707L..42K} is the filter we use to suppress the CMB signal in the WMAP maps. We use a multi-frequency matched filter that optimizes detection of the cluster signal. As well as providing a higher signal to noise measurement, our filter automatically accounts for CMB and noise correlations between the channels. Using simulations we find that if no attempt is made to reduce the thermal SZ signal in the maps, our results would be biased. We use a modified version of the matched filter that incorporates the thermal SZ spectrum, to reduce the thermal SZ signal by over an order of magnitude. In addition, we subtract the monopole and dipole from the region outside of the galactic mask, and use an identical pipeline for the data and error analysis.

We also measure the temperature monopole at the locations of galaxy clusters. The cluster monopole contains a contribution from the thermal and kinetic SZ effects. The kinetic SZ contribution is expected to be zero if there is no cluster monopole velocity, as we expect. In addition we measure the quadrupole, octupole, and $\ell \! = \! 4$ and 5 modes. Our method can be applied to data from the upcoming Planck experiment to achieve a result with greater sensitivity~(Mak et al. 2010).

In section~\ref{sec:theory} we present all necessary theoretical background information. We describe the SZ effect and how it can be used to measure the bulk flow, and present the $\Lambda$CDM predictions for the velocity of the flow. In section~\ref{sec:data} we describe the data that we use to make our measurement, and explain how we construct our cluster sample. In section~\ref{sec:method} we explain our method, describing how we model the SZ emission, how we construct filters to optimize cluster detection, and how we determine the amplitude and direction of any bulk flow. In section~\ref{sec:sys} we describe the level of contamination of thermal SZ, unresolved radio point sources, and galactic emission, which we determine using simulations. In section~\ref{sec:results} we present our results and in section~\ref{sec:conclusions} give our conclusions.

Throughout this paper we assume a spatially flat $\Lambda$CDM cosmology with the WMAP~7 year cosmological parameters~\citep{2010arXiv1001.4744J}.

\section{Theory}
\label{sec:theory}

\subsection{SZ Clusters as Tracers of the Velocity Field}

The SZ effect can be used to measure the deviations of galaxy clusters from the Hubble flow~\citep{1980MNRAS.190..413S}. Approximately $1\%$ of CMB photons traveling through galaxy clusters are scattered by electrons trapped in the gravitational potential of the cluster. If the galaxy cluster is moving with respect to the CMB rest frame, the scattered photons are red or blue shifted. This is the kinetic Sunyaev-Zeldovich (kSZ) effect~\citep{1980MNRAS.190..413S} and results in a fractional temperature change in the radiation of

\begin{equation}
\frac{\Delta T_{\rm KSZ}}{T} = - \int \textbf{dl} \cdot \frac{\textbf{v}_p}{c} n_e \sigma_T
\label{eqn:ksz}
\end{equation}

\noindent where $\textbf{dl}$ is the line of sight distance through the cluster, $\textbf{v}_p$ is the cluster peculiar velocity, $n_e$ is the electron density and $\sigma_T$ is the Thompson scattering cross section. The quantity $\int n_e \sigma_T dl$ is the optical depth to Thompson scattering, $\tau$, with typical value $\sim \! 0.01$ for the clusters we observe. By observing the temperature increment or decrement in the direction of a galaxy cluster, the line of sight cluster peculiar velocity can be estimated. For clusters in our sample we expect $\Delta T_{\rm KSZ}/T \sim 10^{-5}$. To date no detection of a cluster velocity has been made using the kSZ effect.

A bulk motion of mass in the universe will cause a dipole pattern in temperature at the cluster positions. The kSZ signal from each cluster that is part of a bulk motion is given by,

\begin{equation}
\frac{\Delta T_{{\rm KSZ}}}{T} = \frac{v_{\rm{bulk}}}{c} \tau \cos \theta
\label{eqn:ksz_bulk}
\end{equation}

\noindent where $\tau$ is the optical depth of the cluster and $\theta$ is the angle between the cluster position and the bulk flow direction. Cluster motions with more structure than monopole and dipole terms can be described by decomposing the cluster velocity field into spherical harmonics.

The electrons within massive clusters have temperatures that are typically a few keV. The thermal motion of these electrons causes an increase in the energy of the scattered photons, which is the thermal Sunyaev-Zeldovich (tSZ) effect~\citep{1970Ap&SS...7....3S}. The fractional temperature change in the radiation in the non-relativistic limit is

\begin{equation}
\frac{\Delta T_{\rm TSZ}}{T} = \left( x \frac{e^x + 1}{e^x - 1} - 4 \right) \int dl \; \frac{k_B T_e}{m_e c^2} n_e \sigma_T
\label{eqn:tsz_sig}
\end{equation}

\noindent where $x = h \nu / k_B T_{\rm{CMB}}$, $\nu$ is the radiation frequency, $dl$ is the line of sight distance through the cluster and $T_e$ is the electron temperature. To determine the importance of relativistic corrections to equation~\ref{eqn:tsz_sig}, we have calculated the thermal SZ spectrum of all 736 clusters in our cluster sample (we describe the cluster modeling we use in section~\ref{sec:sims}), using equation~\ref{eqn:tsz_sig} to calculate the tSZ emission and also including relativistic corrections using the model of~\citep{2000ApJ...536...31N}. We combine the spectra, weighting the spectrum from each cluster with the same weight that we use when we measure the bulk flow signal (we describe how we calculate the bulk flow signal in section~\ref{sec:analysis}). We find that the difference between the flux calculated using equation~\ref{eqn:tsz_sig} and the flux from the relativistic formula is less than $5\%$ at all of the frequencies we use, and so for our purposes we can safely ignore the relativistic corrections.

For typical clusters $\Delta T_{\rm TSZ}/T \sim 10^{-4}$ at 90 GHz, and the ratio of the kinetic signal to the thermal signal is (assuming the clusters are isothermal),

\begin{equation}
\frac{\Delta T_{\rm KSZ}}{\Delta T_{\rm TSZ}} \sim \frac{v_p m_e c}{k_B T_e } \sim 0.1 \left( \frac{v_p}{300 \rm km/s} \right) \left( \frac{T_e}{5 \rm keV} \right)^{-1}
\end{equation}

The thermal signal is therefore a significant contaminant to any measurement of the kinetic signal. However, since there are not expected to be intrinsic large scale moments in the tSZ signal, the tSZ dipole amplitude is expected to be smaller than the monopole amplitude.

Unlike the tSZ signal, the kSZ signal has an identical frequency spectrum to the CMB. The kSZ signal can therefore only be distinguished from CMB fluctuations by the different spatial properties of the two signals. The CMB power peaks on degree scales, whereas the SZ emission from galaxy clusters typically varies on arcminute scales. The signal we detect is smeared out by the instrumental beams of the experiment, $30^{\prime}.6$ in the WMAP Q band, $21^{\prime}$ in the V band and $13^{\prime}.2$ in the W band.

\subsection{Expected Signal For a Generic Velocity Tracer}
\label{sec:lcdm}

On scales larger than $\sim \! 10 h^{-1}$ Mpc, fluctuations in the matter density of the universe are Gaussian distributed~\citep[][and references therein]{2005MNRAS.360L..82R}. The inhomogeneities in the matter density cause galaxy clusters to have peculiar velocities, with typical values of $\sim \! 300$ km/s~\citep[e.g.,][]{2008PhRvD..77h3004B}. If a sample of galaxy clusters within a region is chosen, that sample will have a non-zero bulk motion due to the non-uniform matter density on even larger scales. Although the direction of this bulk motion is not determined by the $\Lambda$CDM model, the rms amplitude can be calculated given a set of cosmological parameters. We now calculate this amplitude.

\begin{figure}
  \centering
  \includegraphics[width=83mm]{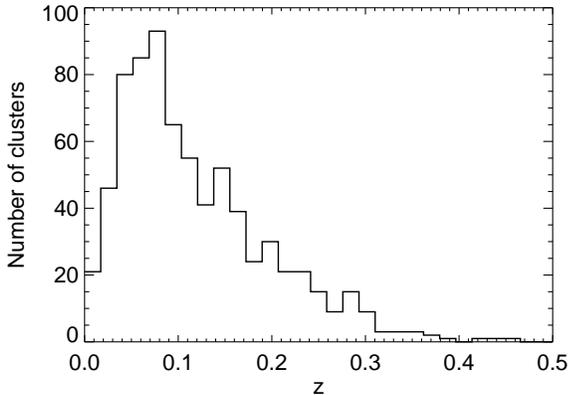}
  \caption{Redshift distribution for the cluster sample used in this paper.}
  \label{fig:zhist}
\end{figure}

We can estimate the expected cluster bulk flow velocity by expanding the line of sight peculiar velocity distribution in spherical harmonics. The rms bulk flow velocity will be the power in the first multipole. We follow the derivation for the density distribution given in~\citet{1973ApJ...185..413P}. We decompose the line of sight peculiar velocity into spherical harmonics, $Y_{\ell m}(\hat{\textbf{r}})$, with the amplitude of the $\ell$,$m$ mode given by:

\begin{equation}
a_{\ell m} = \int dr r^2 \phi(r) \int d \Omega_r Y^{\ast}_{\ell m}(\hat{\textbf{r}}) \; \textbf{v}(\textbf{r}) \cdot \hat{\textbf{r}}
\label{eqn:exp_alm}
\end{equation}

\noindent where $\textbf{r}$ is the comoving radial distance, $\phi$ is the comoving number density of objects in the sample, $\textbf{v}$ is the object peculiar velocity and $\Omega_r$ is the solid angle. We approximate $\phi(r)$ by an isotropic function, which we estimate by calculating the number density of clusters in our sample within radius $r$. We normalize $\phi$ such that:

\begin{equation}
\int_0^R dr r^2 \phi(r) = 1
\end{equation}

where $R$ is the comoving distance within which the $a_{\ell m}$ are calculated. Figure~\ref{fig:zhist} shows the redshift distribution of our cluster sample used to calculate $\phi$.

\begin{figure}
  \centering
  \includegraphics[width=80mm]{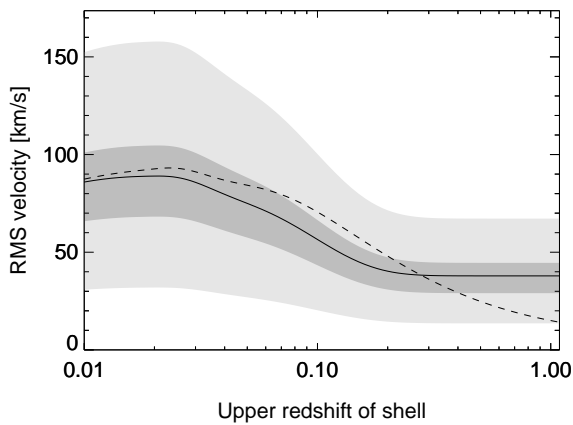}
  \caption{Expected bulk flow velocity in $\Lambda$CDM cosmology with a selection function calculated from our cluster sample in redshift shells extending from $z=0$ to the specified x-axis value. The dashed line shows the result with a uniform selection function. The dark and light shaded areas are the $68\%$ and $95\%$ confidence limits from cosmic variance.}
  \label{fig:theory}
\end{figure}

The power in multipole $\ell$ is defined as $C_{\ell} = \langle | a_{\ell m} |^2 \rangle$ where the average is over all values of $m$ within multipole $\ell$. If the line of sight peculiar velocity over the whole sky is a Gaussian random field, as expected in the $\Lambda$CDM model, then it is completely described by $C_{\ell}$. The power in the dipole ($\ell=1$) is (the details of the derivation are given in appendix~\ref{sec:dipolederiv}):

\begin{equation}
\begin{split}
C_1 &= \langle | a_{1m} |^2 \rangle = \frac{2}{9 \pi} f^2 H_0^2 \int dk P(k) \\
&\left( \int dr r^2 \phi(r) \left( j_1 (kr) - 2j_2 (kr) \right) \right)^2 \\
\label{eqn:exp_bf}
\end{split}
\end{equation}

\noindent where $P(k)$ is the matter power spectrum, $f = (a/D) \; dD/da$, $a$ is the scale factor, $D$ is the growth function, $H_0$ is the Hubble constant and $j_1$ and $j_2$ are spherical Bessel functions. Since the clusters we observe are all at redshift less than one, we approximate $f$ by its value today, taking it to be equal to $\Omega_m^{0.6}$. We obtain a prediction of the rms dipole velocity shown in figure~\ref{fig:theory}. Since our cluster sample contains few clusters with redshifts greater than $\sim \! 0.3$, the bulk velocity of our entire cluster sample is largely determined by clusters with redshifts less than this. The effect of the selection function is therefore to increase the expected dipole velocity in shells extending to redshifts greater than 0.3. The shaded area in figure~\ref{fig:theory} is the uncertainty from cosmic variance. The additional uncertainty on $C_1$ from sample variance is inversely proportional to the number of clusters that are used to measure the dipole. For redshift shells extending beyond $z= 0.05$ our cluster sample contains over one hundred clusters and so the sample variance in figure~\ref{fig:theory} will be more than ten times smaller than the cosmic variance. We therefore do not include it in the figure.

A measurement of a kSZ bulk flow with an amplitude larger than that expected from cosmic variance would be an important result if confirmed, requiring modifications of inflation to explain it~\citep[e.g.,][]{1991PhRvD..44.3737T,1994PhRvL..73.1582K}. A measurement of a kSZ monopole with a velocity greater than that expected by sample variance would be a violation of the Copernican principle, which states that we do not live in a specially favored place in the universe. However, a measurement of the kSZ monopole is more susceptible to contamination. The thermal SZ signal, as well as the radio point source signal from clusters, are both expected to be isotropic, with any dipole signal coming from sample variance. Any dipole signal from tSZ or radio point sources is therefore expected to be weaker than any corresponding monopole signal. The radio source monopole at the cluster locations could be removed by observing and subtracting the emission, or by utilizing the different spectra of the radio source and kSZ signals. For thermal SZ the monopole can be suppressed spectrally.

\section{Data}
\label{sec:data}

We have searched for a cluster dipole in the WMAP~7 year maps\footnote{The WMAP maps, beam functions and galactic masks are available at \url{http://lambda.gsfc.nasa.gov/product/map/current/m_products.cfm}}~\citep{2010arXiv1001.4744J}. We use the foreground reduced maps produced by WMAP. These were generated by removing a foreground model from each `unreduced' map~\citep{2007ApJS..170..288H}. Both the K and Ka band maps were used to produce the galactic foreground model, and so there are no K or Ka band foreground reduced maps. The maps are pixelized in the Healpix format~\citep{2005ApJ...622..759G} with a pixel size of $7^{\prime}$. The CMB dipole has been removed from each map. Removing the CMB dipole from the maps does not significantly change the cluster dipole that we search for. We have tested the effect of removing the CMB dipole by simulating kSZ maps (our simulation procedure is described in section~\ref{sec:sims}), adding a CMB dipole signal, and then removing it using a least squares fit to all map pixels. We find that the amplitude of the cluster dipole in the resulting Q band maps is changed by less than $1\%$, and the direction of the dipole is changed by less than $25^{\prime}$. In the V and W band channels the change is smaller due to a stronger cluster signal relative to the CMB dipole signal.

We use the two Q (41 GHz), two V (61 GHz) and four W (94 GHz) band maps from each differencing assembly. We exclude the K and Ka band maps from the analysis since no foreground reduced maps are available. The galactic synchrotron and extra-galactic radio emission has a spectral index $\alpha > 0$ (with source flux, $S \sim \nu^{-\alpha}$), and the emission is stronger in the K and Ka band than in the Q, V or W bands. Although the instrument noise is lower in the K and Ka bands than in the Q, V or W bands, the cluster signal is weaker, due to dilution by the larger beam FWHM. Including the K and Ka band channels therefore only increases the signal to noise of the cluster dipole measurement by $\sim \! 10\%$ and so we exclude them from our analysis.

The noise per pixel is $\sim \! 65 \mu$K (Q band), $\sim \! 80 \mu$K (V band), and $\sim \! 137 \mu$K (W band). The noise is lower near the ecliptic poles due to the WMAP scan strategy. The beam FWHM are $30^{\prime}.6$ (Q band), $21^{\prime}$ (V band), and $13^{\prime}.2$ (W band). We generate noise simulations using the prescription described in~\citet{wmap_3year_suppl}. The noise is assumed to be uncorrelated between pixels, which is a good approximation for the temperature maps that we use~\citep{wmap_3year_suppl}. We use the WMAP beam transfer functions, which are the square roots of the window functions used for the power spectrum analysis, to construct our matched filters and smooth our simulated maps. The WMAP extended temperature analysis mask is used to exclude the galaxy and other known regions of high foreground emission. Any cluster that lies behind this mask is not included in our cluster sample.

\subsection{The X-ray Selected Cluster Sample}
\label{sec:xraysamp}

Following KAKE we derive our cluster sample from the REFLEX~\citep{2004A&A...425..367B}, BCS~\citep{2000yCat..73010881E}, Extended BCS~\citep{2000yCat..73180333E} and CIZA~\citep[\citealp{2002ApJ...580..774E},][]{2007ApJ...662..224K} cluster catalogs. These catalogs contain 447, 206, 107, and 130 clusters respectively. After removing 20 overlapping clusters we obtain a sample of 870 clusters. After removing clusters whose center lies behind the WMAP galactic mask we find 736 clusters (when using the WMAP Kp0 galactic mask that was used in the KAKE analysis we find 771 clusters in the sample). The redshift distribution of the cluster sample is shown in figure~\ref{fig:zhist}. We convert the luminosities in the BCS, BCSe and CIZA catalogs to the values for a $\Lambda$CDM universe. For the REFLEX sample we use a catalog with the luminosities already converted to an $h = 0.7$, $\Lambda$CDM cosmology so no correction is necessary for these clusters.

To derive average electron density and temperature profiles for our cluster sample we use the Archive of Chandra Cluster Entropy Profile Tables~\citep[ACCEPT,][]{2006ApJ...643..730D,2009ApJS..182...12C}. This is a sample of 239 clusters observed by the Chandra X-ray Observatory~\citep{2000SPIE.4012....2W}, with electron density and temperature profiles for each cluster derived from a deprojection analysis. 145 of the ACCEPT clusters overlap with our sample.

\subsection{External Simulations}
\label{sec:extsim}

To test for contamination from galactic emission, thermal SZ and unresolved radio sources, we use simulated maps produced using the Planck Sky Model\footnote{\url{http://www.apc.univ-paris7.fr/APC_CS/Recherche/Adamis/PSM/psky-en.php}} (PSM).

The PSM is a set of programs and data used for the simulation of full sky microwave maps that includes galactic synchrotron, dust and free-free emission, kinetic and thermal SZ, radio and infrared point sources and CMB. The PSM simulations do not include a kinetic SZ monopole or cluster bulk flow velocity. We use one realization of the sky generated at the WMAP Q, V and W band center frequencies.

To test the effect of radio point sources we use 100 simulated maps of the unresolved radio background produced by~\citet{2010arXiv1001.3659C}. They combine sources from the NRAO-VLA Sky Survey (NVSS) catalog~\citep{1998AJ....115.1693C} with higher frequency surveys to extrapolate the radio point source emission at $1.4$ GHz to the WMAP frequencies. The radio source flux distribution in the maps at each frequency therefore reflects the probability distribution for the flux of each NVSS source.

\section{Method}
\label{sec:method}

\subsection{Outline of the Method}

The bulk motion of many galaxy clusters relative to the CMB rest frame creates a signal in the WMAP temperature maps that has a dipole pattern at the location of clusters. We now describe the method we use to recover the amplitude and direction of any dipole motion using the WMAP CMB maps.

In addition to kSZ the dipole signal at the position of clusters includes contributions from CMB, instrumental noise, tSZ, galactic emission and infrared and radio point source emission. The first step is to filter the maps to enhance the kSZ signal relative to the other terms. We use two filters, one of which enhances the cluster signal relative to the CMB and instrument noise, the other filter also removes the thermal SZ term. The galactic term is suppressed by masking the WMAP maps with the WMAP extended temperature analysis mask, and only fitting for a cluster dipole outside of this region. In section~\ref{sec:sys_radio} we give upper limits on the radio point source contamination and in section~\ref{sec:gal} we give upper limits on the galactic signal. Using the PSM simulations we find a negligible contribution from infrared point source emission.

The next step is to calculate the monopole and dipole at the locations of the ROSAT observed clusters in the filtered maps. We use simulated kinetic SZ maps to convert the amplitude of the dipole at the cluster locations to a bulk flow velocity using a method we describe in section~\ref{sec:conv_dip}. Our simulated kinetic and thermal SZ maps include signal only from the ROSAT observed clusters that we use, and will be described in section~\ref{sec:sims}.

There are two ways we could calculate the cluster velocity dipole. We could calculate the optical depth through each cluster and use equation~\ref{eqn:ksz} to estimate the velocity of each cluster. We could then fit for a dipole in the cluster velocity distribution. Alternatively, we could calculate the dipole at the cluster positions in the WMAP maps and convert the dipole signal to a velocity using an estimate of the average optical depth of the clusters in our sample. The two methods differ in the weight given to each cluster in the dipole fit. We expect the signal to noise of the latter method to be higher since all clusters are given equal weight in the fit, whereas in the former method clusters with a larger optical depth (and hence a greater kSZ signal in the map) are given less weight. We follow KAKE and choose the latter method. We split the cluster sample into different redshift bins (some of which overlap) and calculate the cluster dipole in each.

To calculate the errors on the monopole, dipole and higher order modes, we generate 100 realizations of the CMB, convolve the maps with the beams from each of the eight channels and add instrument noise using the procedure described in~\citet{wmap_3year_suppl}. We pass these maps through our pipeline to find the distribution of monopole, dipole and higher order modes.

\subsection{Cluster modeling}
\label{sec:sims}

Our measurement of the WMAP temperature dipole does not depend on any cluster simulations. However, in order to convert the cluster temperature dipole measurement into a velocity dipole we require simulated maps of the kSZ emission from our cluster sample. In addition, we require tSZ realizations to estimate the level of tSZ contamination in our results.

The clusters in our sample have typical angular sizes of $\sim \! 1^{\prime}$, which is small compared with the WMAP beam FWHM, $30^{\prime}.6$ in Q band and $13^{\prime}.2$ in W band. For this reason we model the clusters as point sources convolved by the WMAP beams.

\subsubsection{Optical Depth Determination}
\label{sec:tau_calc}

In order to generate kSZ realizations with a simulated bulk flow velocity we require the optical depth to Thompson scattering for every cluster in our sample. We calculate this using average electron density and temperature profiles that we derive from the ACCEPT catalog (described in section~\ref{sec:xraysamp}).

To calculate the electron density and temperature profiles we select 145 ACCEPT clusters that overlap with our sample. We average the electron density and temperature profiles of the overlapping clusters within radial bins. We use bins with a smaller size nearer to the center of the cluster where the error on the electron density and temperature is lower (the error is provided in the catalog). The mean bin size is 10 kpc, and the largest radial bin extends to 1 Mpc from the cluster center. Each profile is normalized so that it has unit value at the center of the cluster. The densities and temperatures of all of the clusters are then averaged within each radial bin. We compute the kSZ and tSZ signals using these profiles, giving the profiles a different normalization for each cluster, as described in the following paragraph. Extrapolating the density profile to larger radii using a $\beta-$model~\citep{1976A&A....49..137C}, we find a negligible contribution to the optical depth from beyond this region. The profiles are not significantly changed when all of the clusters in the ACCEPT sample are used to calculate them.

Since we use the same electron density profile for each cluster, we only need to calculate the normalization to that profile to estimate the optical depth. The optical depth is calculated by integrating the electron density along the line of sight,

\begin{equation}
\tau(R) = \sigma_T N_e \int \frac{f_e(r) r dr}{\sqrt{r^2-R^2}}
\end{equation}

\noindent where $R = d_A \theta$ is the distance from the cluster center perpendicular to the line of sight, $d_A$ is the angular diameter distance, $\sigma_T$ is the Thompson scattering cross section, $f_e(r)$ is the normalized electron density profile and $N_e$ is the normalization to that profile.

We calculate the normalization to the electron density profile by requiring that the cluster bolometric luminosity in the ROSAT catalogs is equal to the luminosity from Bremsstrahlung emission, which depends on $N_e$~\citep{1978A&AS...32..283G},

\begin{equation}
\begin{split}
L = &\frac{32 \pi}{3} \left( \frac{2\pi k_B}{3 m_e} \right)^{\frac{1}{2}} \frac{Q^6}{m_e c^3 h} Z^2 g N_e N_i \\
&\int d^2 r \; 4 \pi r^2 \; f_e (r) f_i (r) T(r)^{\frac{1}{2}} \;\; \rm{erg/s}
\label{eqn:Lbol}
\end{split}
\end{equation}

\noindent where $k_B$ is the Boltzmann constant, $m_e$ is the electron mass, $Q$ is the electron charge, $c$ is the speed of light, $h$ is Planck's constant, Z is the number of protons per nucleus, $g$ is the frequency averaged Gaunt factor (of order unity), $T(r)$ is the electron temperature, $f_e(r)$ and $f_i(r)$ are the normalized electron and ion density profiles (which we take to be equal) and $N_e$ and $N_i$ are the normalizations to those profiles. We normalize the temperature profile such that the X-ray emission weighted temperature is equal to $T_X$, which we calculate from the X-ray luminosity of each cluster using the relation $T_X = (2.76 \pm 0.08)L_X^{0.33 \pm 0.01}$~\citep{1997MNRAS.292..419W}.

Using this procedure we find that the mean central optical depth of our cluster sample is $(4.9 \pm 0.9) \times 10^{-3}$. The mean central optical depth of the 145 clusters that overlap with the ACCEPT sample is $\tau_0 = (6.1 \pm 1.0) \times 10^{-3}$ which is in agreement with the value calculated using each individual electron density profile in the ACCEPT catalog, $\tau_0 = (6.7 \pm 0.9) \times 10^{-3}$. This gives us confidence that we can calculate accurate values of the optical depth for clusters that are not in the ACCEPT sample. The dominant source of uncertainty in our optical depth calculation is the cluster X-ray luminosity, from the X-ray catalogs. Propagating all errors through to the optical depth gives uncertainties of $\sim \! 15\%$. These errors are propagated through to the simulated kSZ maps which allows us to calculate the precision of our method.

We have not accounted for any anisotropy in the optical depth distribution of our clusters or accounted for the different methods used to calculate the X-ray luminosities in the different X-ray catalogs we use. Both of these steps would require an independent analysis of the X-ray data. However, since our dipole fit does not depend on the X-ray luminosities our null detection is unaffected. Any error in the X-ray luminosities will only affect the conversion of the dipole amplitude from $\mu$K to km/s. Since the X-ray luminosities are used to calculate the optical depth any asymmetry in the optical depths will be accounted for when converting from $\mu$K to km/s.

\subsubsection{kSZ and tSZ Simulations}

The kSZ signal is calculated from equation~\ref{eqn:ksz}. We give each cluster in our simulated maps a bulk velocity using equation~\ref{eqn:ksz_bulk} and a random velocity drawn from a Gaussian distribution, with variance~\citep[e.g.,][]{1988ApJ...332L...7G}

\begin{equation}
\label{eqn:sigv}
\sigma^2_{\rm{v}} = \frac{f^2 H_0^2}{6\pi^2} \int_0^{\infty} d k \; P(k) \; | W(kR) |^2
\end{equation}

\noindent where $| W(kR) |^2$ is the Fourier transform of the window function, and again we use $f \approx \Omega_m^{0.6}$. To select cluster scales we use a top hat window function with $R = 10 h^{-1}$ Mpc, and find $\sigma_{\rm{v}} = 272$ km/s. We ignore correlations between the velocities of different clusters, which we expect to be small for our cluster sample~\citep[eg.][]{1988ApJ...332L...7G, 2008PhRvD..77h3004B}.

The expected tSZ signal is calculated by integrating equation~\ref{eqn:tsz_sig} through the cluster using the same electron density and temperature profiles calculated in section~\ref{sec:tau_calc}. Since WMAP is sensitive enough to detect the tSZ signal in stacked images we have checked that the average tSZ signal in our simulations agrees with WMAP observations. Following the method in~\citet{2008ApJ...675L..57A} (also \citealp{2010MNRAS.402.1179D}), we have stacked maps centered on each cluster to produce a coadded map in the Q, V and W bands from the WMAP maps and our simulated tSZ maps. We find that the residual after subtracting the two is consistent with CMB and instrument noise.

In table~\ref{tbl:tsz} we show the simulated thermal SZ monopole in maps filtered by the matched filter we use in our analysis. We compare this to the monopole found in the WMAP maps filtered by the same filter. Although the monopole in our simulated maps is biased high it is within $2 \sigma$ of the measured monopole in all except the two lowest redshift shells. This gives us confidence that our optical depth calculation is reasonable.

\begin{deluxetable}{ccccc}
\tabletypesize{\footnotesize}
\tablecolumns{5}
\tablewidth{0pc}
\tablecaption{Simulated thermal SZ monopole and measured monopole.\label{tbl:tsz}}
\tablehead{
	\colhead{$z_{min}$}						&
	\colhead{$z_{max}$}						&
	\colhead{$N_{\rm{cl}}$}						&
	\colhead{Simulated tSZ Monopole, $a_0$ [$\mu$K]}		&
	\colhead{Measured Monopole, $a_0$ [$\mu$K]}			\\
}
\startdata
0.0  & 0.04 & 95  & -78.7  & -22  $\pm$ 33  \\
0.0  & 0.05 & 139 & -84.2  & -41  $\pm$ 27  \\
0.0  & 0.06 & 192 & -97.0  & -62  $\pm$ 24  \\
0.0  & 0.08 & 294 & -97.9  & -63  $\pm$ 18  \\
0.0  & 0.12 & 445 & -99.5  & -73  $\pm$ 13  \\
0.0  & 0.16 & 546 & -103.4 & -81  $\pm$ 12  \\
0.0  & 0.20 & 619 & -109.1 & -85  $\pm$ 12  \\
0.0  & 1.0  & 736 & -123.9 & -91  $\pm$ 11  \\
0.05 & 0.30 & 578 & -128.6 & -99  $\pm$ 12  \\
0.12 & 0.30 & 271 & -152.2 & -112 $\pm$ 17  \\
\enddata
\end{deluxetable}

\subsection{Filters}
\label{sec:filters}

The distribution of our cluster sample on the sky is not isotropic and so measurements of the peculiar velocity multipoles will be contaminated by higher order CMB multipoles. The higher order CMB multipoles couple to the signal we are trying to measure through the mask we use to cover non-cluster regions of the WMAP maps. To reduce the contamination from the CMB we filter the maps.

Two filters are used, and described here. The first is an optimal matched filter that removes low-order CMB multipoles by utilizing information at locations other than the cluster position to distinguish a CMB multipole from a kSZ multipole. The second filter uses both spatial information and the spectral difference between the thermal and kinematic SZ effects to remove the contaminating tSZ signal.

An optimal filter is constructed for each channel such that when the filtered maps from each channel are combined, the kSZ signal to noise ratio is maximized. In order to construct our filters we require a matrix of cross spectra, $\textbf{C}_{\ell}=\textbf{C}^{\rm CMB}_{\ell} + \textbf{C}^{\rm noise}_{\ell}$, that describes the statistical properties of the maps to be filtered. We assume that the noise in map $\nu$, $n_\nu (\theta)$, is a homogeneous and isotropic random field. The cross power spectrum of maps $\nu_1$ and $\nu_2$ is equal to $\left\langle n_{\ell m,\nu_1}n_{\ell m,\nu_2}^\ast \right\rangle$, where $n_{\ell m,\nu}$ are the coefficients in the spherical harmonic expansion of $n_\nu (\theta)$, and the average is over $m$. We generate $\textbf{C}_{\ell}$ by averaging the cross power spectra of 100 CMB and noise realizations.

The maps in each channel are filtered and combined to produce the spherical harmonic coefficients of the filtered map as follows,

\begin{equation}
f_{\ell m} = \sum_{i=1}^{n_{\rm{channels}}} \Phi^i_{\ell} a^i_{\ell m}
\end{equation}

\noindent where $a^i_{\ell m}$ are the spherical harmonic coefficients of the WMAP channel $i$ map and $\Phi^i_{\ell}$ is the filter for channel $i$. We include the factor of $\sqrt{4\pi/2\ell+1}$ from spherical convolution in $\Phi_{\ell}$. The first filter we use has a filter function of the form~\citep{2008ITSP...56.3813M}:

\begin{equation}
\Phi^{\rm{m}}_{\ell} = \frac{\textbf{C}_{\ell}^{-1}}{\gamma} \textbf{B}_{\ell}
\end{equation}

\noindent where m indicates that the filter is a matched filter, $\gamma = (1/n_{\rm{pix}})\sum_{\ell} (2\ell+1) \textbf{B}^T_{\ell} \textbf{C}_{\ell}^{-1} \textbf{B}_{\ell}$, $n_{\rm{pix}}$ is the number of map pixels and $(4\pi/n_{\rm{pix}}^2) \textbf{B}_{\ell}^2$ is an estimate of the kSZ power spectrum from a single cluster, in each channel (we assume the beam convolved sources are symmetric). Since the WMAP beams are much larger than galaxy clusters we approximate $\textbf{B}_{\ell}$ by the WMAP beam function. We have calculated the filter using the WMAP beam profile convolved with an average cluster profile and also a cluster profile modeled as a $\beta-$model~\citep{1976A&A....49..137C} and find no significant difference in the results.

A simple modification to this filter allows the thermal SZ bias to be removed~\citep[e.g.,][for the flat and curved sky cases respectively]{2005MNRAS.356..944H,2006MNRAS.370.1713S},

\begin{equation}
\Phi^{\rm{u}}_{\ell} = \frac{\textbf{C}_{\ell}^{-1}}{\Delta} \left( \alpha \textbf{B}_{\ell} - \beta \textbf{F}_{\ell} \right)
\label{eqn:sz_rem_filt}
\end{equation}

\noindent where u indicates that the filter is an unbiased filter, $\textbf{F}_{\ell}$ is an estimate of the thermal SZ signal in each channel, $\Delta = \alpha \gamma - \beta^2$ is a normalization factor, $\alpha = (1/n_{\rm{pix}})\sum_{\ell} (2\ell+1) \textbf{F}^T_{\ell} \textbf{C}_{\ell}^{-1} \textbf{F}_{\ell}$, $\beta = (1/n_{\rm{pix}})\sum_{\ell} (2\ell+1) \textbf{B}^T_{\ell} \textbf{C}_{\ell}^{-1} \textbf{F}_{\ell}$ and again $\gamma = (1/n_{\rm{pix}})\sum_{\ell} (2\ell+1) \textbf{B}^T_{\ell} \textbf{C}_{\ell}^{-1} \textbf{B}_{\ell}$. We are using maps with units of thermodynamic temperature and so $\textbf{F}_{\ell} = \left[x (e^x+1)/(e^x-1) - 4.0 \right] \textbf{B}_{\ell}$.

\begin{figure}
  \centering
  \includegraphics[width=82mm]{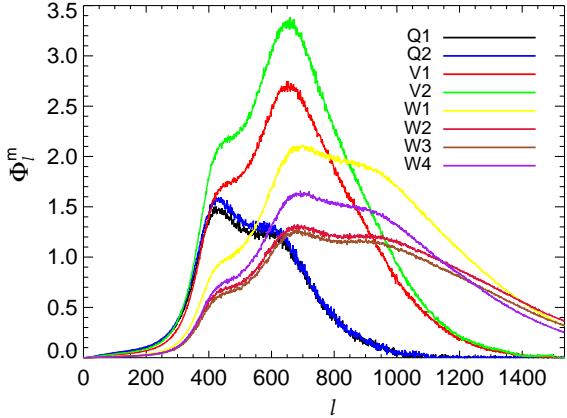}
  \caption{Multi-frequency matched filters for the WMAP channels.}
  \label{fig:matched_filter1}
\end{figure}

\begin{figure}
  \centering
  \includegraphics[width=82mm]{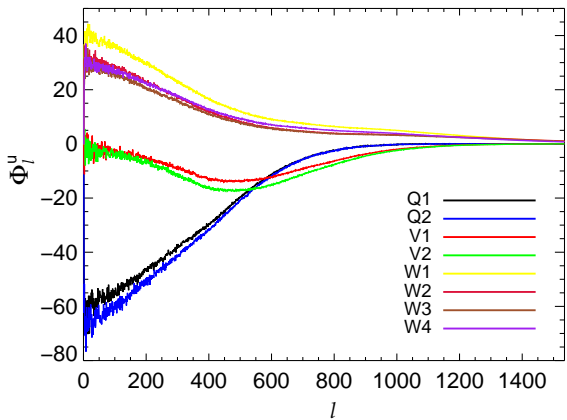}
  \caption{Unbiased multi-frequency matched filters. A discussion of the filter shape is given in the text.}
  \label{fig:matched_filter2}
\end{figure}

Figures~\ref{fig:matched_filter1} and~\ref{fig:matched_filter2} show the functions $\Phi^{\rm{m}}_{\ell}$ and $\Phi^{\rm{u}}_{\ell}$ for each channel. These filters have very different behaviors and we now discuss why this arises.

The matched filter (hereafter MF) suppresses the input map at multipoles less than $\sim \! 300$ (large scales) due to the strong CMB signal at these multipoles. The filter function also suppresses the map at multipoles greater than $\sim \! 1000$ (small scales), where the expected kSZ signal to noise ratio is low. The W band filter is non-zero up to the highest multipoles since the W band beams are smaller, and so the signal extends to higher multipoles. Between multipoles 300-1000 the signal to noise ratio is highest and so these scales are retained in the filtered map. The bumps near multipoles 500 and 700 in the MF correspond to the troughs in the CMB spectrum$-$the filter amplifies scales where the contaminating CMB emission is weaker.

\begin{figure}
  \centering
  \includegraphics[width=80mm]{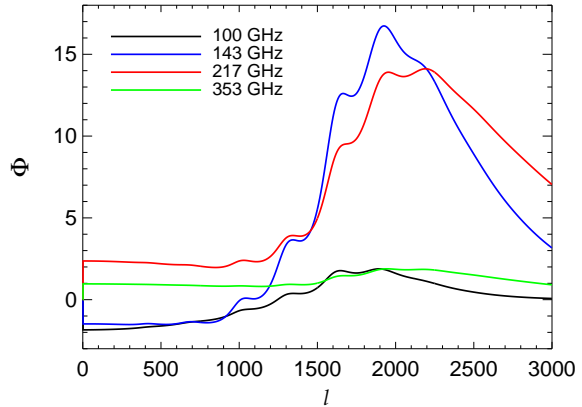}
  \caption{Thermal SZ bias removing filters for an experiment with four frequencies: 100 GHz, 143 GHz, 217 GHz and 353 GHz, beam FWHM of $10^{\prime}$, $7^{\prime}$, $5^{\prime}$ and $5^{\prime}$, and white noise levels 25, 15, 25 and 75 $\mu$K$/$K$ - \rm{arcmin}$.}
  \label{fig:other_filters}
\end{figure}

The tSZ bias removing filter (hereafter UF) in figure~\ref{fig:matched_filter2} is less intuitive. At low multipoles the CMB signal in each channel is almost identical because each channel observes the same CMB sky, and at low multipoles the beam functions are all close to unity. At low multipoles the CMB can therefore be removed by subtracting the Q and W band maps. Since there are two Q band channels and four W band channels the CMB is removed by giving the Q band maps double the weight of the W band maps in the subtraction. This explains why the two Q band filters have an absolute value double that of the four W band filters at low multipoles. This is reflected in equation~\ref{eqn:sz_rem_filt} by the off-diagonal elements being almost equal to the diagonal elements in the cross-power spectra matrix at low multipoles. If the CMB sky in each channel were different then the filter would look similar to the MF but with different amplitudes for each channel. In figure~\ref{fig:other_filters} we show what the tSZ bias removing filters would look like for an  experiment with four frequency channels: 100 GHz, 143 GHz, 217 GHz and 353 GHz, with beam FWHM of $10^{\prime}$, $7^{\prime}$, $5^{\prime}$ and $5^{\prime}$ respectively, and white noise levels: 25, 15, 25 and 75 $\mu$K$/$K$ - \rm{arcmin}$. The fact that there are no peaks in the WMAP filters is a reflection of the limited frequency coverage and large beams.

Since the UF combines channels in a way that removes the tSZ signal, the filters do not increase the kSZ signal to noise ratio as much as the MF. The signal to noise of the cluster temperature dipole measurement is therefore lower.

\begin{figure}
  \centering
  \includegraphics[width=80mm]{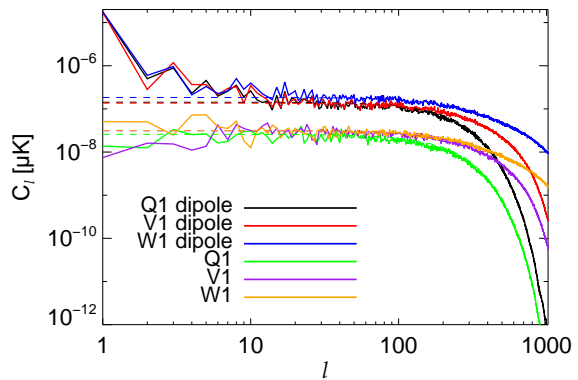}
  \caption{Spectra of a simulated kSZ map (with signal only at the locations of our cluster sample) convolved with the WMAP beams. The simulations were performed both with (top 3 spectra) and without (bottom 3 spectra) a bulk flow component of amplitude 1000 km/s. The dashed lines are the WMAP beam functions, scaled to the signal amplitude.}
  \label{fig:ftest}
\end{figure}

Figure~\ref{fig:ftest} shows the spectra of a simulated map containing only the kSZ signal at the ROSAT cluster locations with and without a bulk flow component of 1000 km/s, and convolved by the WMAP beams. The maps with and without a bulk flow both have spectra that are identical to the WMAP beam function at high multipoles. The map with the bulk flow component contains more power at all multipoles as well as additional power at the dipole and octupole modes. Although the difference between the spectra with and without a bulk flow looks predominantly like an amplitude shift, the bulk flow signal is recoverable by information contained in the spherical harmonic $m$ modes, which are averaged to create the spectra. Our filters are designed to optimize the detection of sources shaped like the WMAP beams. The similarity of the simulated kSZ spectrum with a bulk flow component to the beam profile indicates that our filter will have the desired effect of increasing the signal to noise of the cluster dipole measurement.

\begin{figure}
  \centering
  \includegraphics[width=80mm]{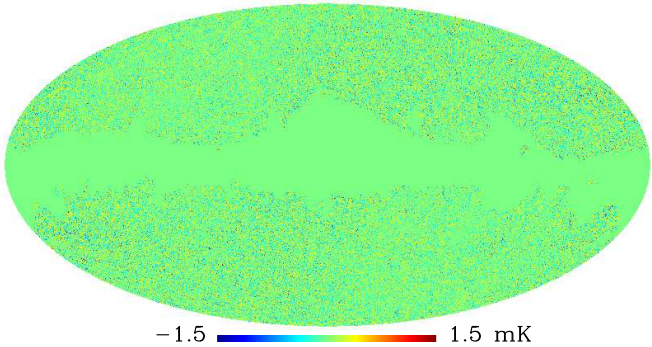}
  \caption{Sum of the eight WMAP maps in the Q, V and W bands filtered by the matched filters in figure~\ref{fig:matched_filter1}. The map is noise dominated.}
  \label{fig:wmapmap_matched_nosz}
\end{figure}

\begin{figure}
  \centering
  \includegraphics[width=80mm]{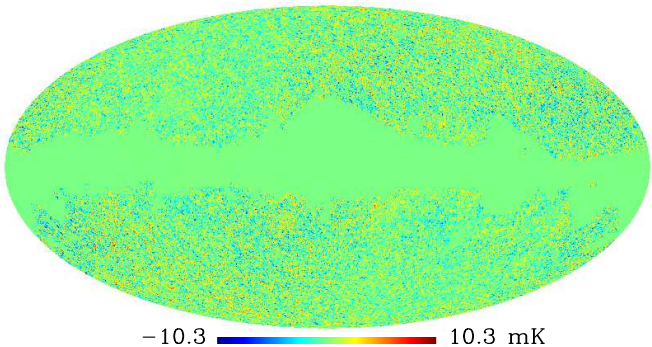}
  \caption{Sum of the eight WMAP maps in the Q, V and W bands filtered by the tSZ removing filters in figure~\ref{fig:matched_filter2}. The values in this map are larger than in figure~\ref{fig:wmapmap_matched_nosz} since the map is noise dominated and the unbiased filters give a lower signal to noise measurement of the kSZ signal.}
  \label{fig:wmapmap_matched}
\end{figure}

Figures~\ref{fig:wmapmap_matched_nosz} and~\ref{fig:wmapmap_matched} show the WMAP maps after filtering with the MF and UF respectively. Both filters have suppressed the CMB signal and the only visible structure in the map is the low noise region around the ecliptic poles. The maps are noise dominated; the larger values of the map in figure~\ref{fig:wmapmap_matched} reflects the lower signal to noise ratio obtained with the UF.

Since we are only calculating the dipole in a small region at the center of the clusters, we could filter the maps only in this region. This would save computation time but care would need to be taken to avoid ringing effects around the edge of the mask. For this reason we filter the maps in all regions outside the WMAP galactic mask.

\subsubsection{Wiener filters}

\begin{figure}
  \centering
  \includegraphics[width=83mm]{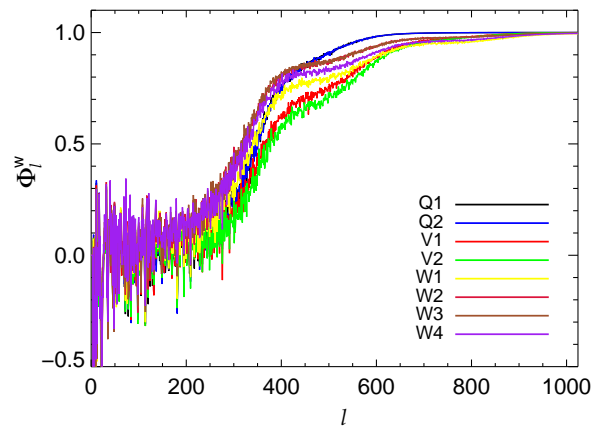}
  \caption{Wiener filter used by KAKE.}
  \label{fig:kf}
\end{figure}

\begin{figure}
  \centering
  \includegraphics[width=83mm]{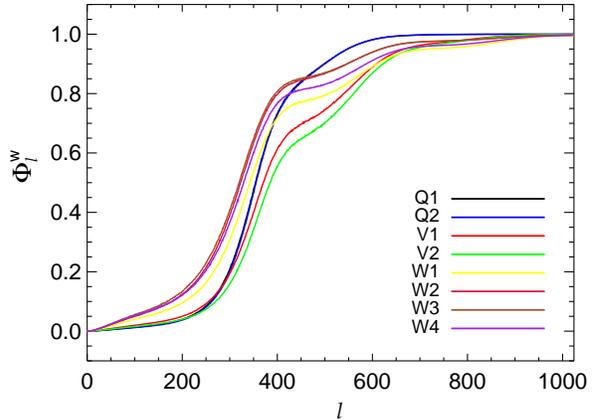}
  \caption{Wiener filter for each of the WMAP channels.}
  \label{fig:wf}
\end{figure}

We now compare our filter to the Wiener filter used by KAKE. A Wiener filter (hereafter WF) minimizes the squared difference between the filtered map and an estimate of the signal. Although a multi-frequency WF can be constructed, we follow KAKE and create a filter for each map that is designed to suppress the CMB component. The KAKE filter is~\citep{2009ApJ...691.1479K}:

\begin{equation}
\Phi^{\rm{w}}_{\ell} = \frac{C_{\ell}^{\rm sky}/f_{\rm sky} - C_{\ell}^{\rm CMB} B_{\ell}^2}{C_{\ell}^{\rm sky}/f_{\rm sky}}
\end{equation}

where $C_{\ell}^{\rm CMB}$ is the CMB power spectrum, $B_{\ell}$ is the WMAP beam function, $C_{\ell}^{\rm sky}$ is the spectrum of the WMAP map estimated outside of the galactic mask and $f_{\rm sky}$ is the fraction of the sky outside of the mask. The full sky spectra could alternatively be calculated by deconvolving the mask: $C_{\ell}^{\rm unmasked} = \sum_{\ell^{\prime}} M^{-1}_{\ell\ell^{\prime}} C_{\ell}^{\rm masked} \approx C_{\ell}^{\rm masked}/f_{\rm sky}$, where $M_{\ell\ell^{\prime}}$ is the multipole mixing matrix which accounts for the cut sky and is calculated from the galactic mask (see eg. appendix A of~\citet{2002ApJ...567....2H} for details). The KAKE filter we use is shown in figure~\ref{fig:kf}. We have determined that this is the same Wiener filter that was used in the KAKE analysis and give details of our pipeline in appendix~\ref{sec:kcomp}. The KAKE filter is constructed from the spectrum of the map that is to be filtered, and maps filtered with it suffer from low multipole fluctuations, caused by cosmic variance. We create our Wiener filter in such a way that it does not suffer from this problem~\citep{1996MNRAS.281.1297T},

\begin{equation}
\label{eqn:wiener}
\Phi^{\rm{w}}_{\ell} = \frac{C_{\ell}^{\rm{noise}}/f_{\rm sky}}{C_{\ell}^{\rm{CMB}} B_{\ell}^2 + C_{\ell}^{\rm{noise}}/f_{\rm sky}}
\end{equation}

\noindent where $C_{\ell}^{\rm{noise}}$ is the power spectrum of the instrument noise plus all foreground components in the masked map. This filter is not normalized and so pixels in the filtered map are not equal to the estimated amplitudes of the kSZ signal in the unfiltered map pixels. Since we model our clusters as point sources convolved by the beam the normalization factor would be $n_{\rm{pix}}/\sum_{\ell} (2\ell+1)B_{\ell}\Phi_{\ell}$. The normalization is instead applied later when the dipole amplitude is converted from Kelvin to km/s. Our Wiener filter is shown in figure~\ref{fig:wf}.

\begin{figure}
  \centering
  \includegraphics[width=85mm]{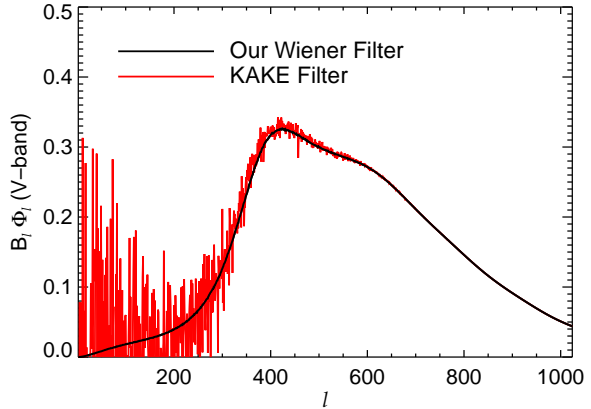}
  \caption{Suppression of the kSZ signal by the beam and filter in the V band.}
  \label{fig:comp_ksz}
\end{figure}

Since the noise in the WMAP maps has a flat spectrum and the CMB signal decreases at high multipoles, the WF acts like a high pass filter for each of the WMAP channels. At low multipoles the shape of this filter is similar to our MF (although the relative normalizations of the channels is not), but at high multipoles the shapes differ since the WF does not take into account the fact that the cluster signal decays exponentially with increasing multipole, while the noise spectrum remains flat. We find that both our Wiener filter and the KAKE filter suppress the cluster signal approximately equally as shown in figure~\ref{fig:comp_ksz}. At the multipoles where the signal peaks ($\sim 400$ in the V band) the two filters suppress power by a similar amount. We find that our Wiener filter reduces the noise in the $a_{1m}$ by a factor of $\sim 2$ relative to the KAKE filter by reducing CMB noise at multipoles less than $\sim 100$ that is correlated between the WMAP channels (we give a more detailed comparison of the filters in appendix~\ref{sec:kcomp}), and is therefore more sensitive to the cluster signal.

For the matched filter and unbiased filter we calculate the cluster dipole in the pixels that overlap with the centers of each cluster. For our Wiener filter we instead use a $15^{\prime}$ aperture around each cluster, in order to increase the signal to noise of the measurement. We have tried fitting for the dipole using different aperture sizes out to $30^{\prime}$ as well as different weighting schemes for the pixels within the aperture: uniform weighting, weighting by the inverse noise variance and weighting by the expected cluster signal to noise ratio. The results presented in table~\ref{tbl:main_WF} have a uniform weighting within the aperture, which was chosen to give a good signal to noise measurement while keeping our results free from any bias caused by incorrect signal or noise estimates.

The WF that we use is not a multi-frequency filter (although in principle a multi-frequency WF could be used) and so there is a filtered map for each of the WMAP channels. We filter each map and compute the cluster dipole in each. We then average the dipoles from the different channels with uniform weights:

\begin{equation}
a_{1m} = \frac{1}{8} \sum_{i=1}^8 a_{1m}^{(i)}
\end{equation}

The signal to noise could be increased by weighting the results from each channel by the expected signal to noise ratio of the kSZ signal. In this scheme the weights given to each channel are approximately equal$-$the noise at higher frequencies is compensated for by the decreased beam FWHM, and therefore larger cluster signal.

\subsection{Dipole Fitting Procedure}
\label{sec:analysis}

To measure the cluster dipole we perform a weighted least squares fit to a dipole function at the locations of the clusters in the filtered map. We also fit for the monopole and, separately, we perform the fit for modes up to $\ell=2$, and up to $\ell=3,4,5$. Our dipole fitting method is based on the Healpix IDL procedure \emph{remove\_dipole}~\citep{2005ApJ...622..759G}. We fit for a vector of coefficients, $\beta$, of the real spherical harmonics using the least squares formula,

\begin{equation}
\beta = (X^T W X)^{-1} X^T W y
\end{equation}

\noindent where $y$ is the data map, $W$ is a diagonal matrix with diagonal values equal to a weights map, and $X$ is a matrix giving the contribution of the fitting function to each pixel. If a monopole and three dipole coefficients are fit to the data, then $X$ is an $n_{\rm{pix}} \times 4$ matrix. We choose the weights map to have non-zero values only in the central cluster pixels, and equal to the inverse noise variance of the filtered map. The noise variance is calculated from 100 filtered WMAP CMB and noise realizations. The vector $\beta$ are the coefficients of the real spherical harmonics, $R_{lm}$, defined by

\begin{equation}
R_{lm} =
\begin{cases}
Y_{{\ell} 0} & \mbox{if $m=0$} \\
{ \frac{1}{ \sqrt{2}} }  \left( Y_{\ell m}+(-1)^m \, Y_{\ell -m} \right) & \mbox{if $m>0$} \\
{ \frac{1}{i\sqrt{2}} } \left( Y_{\ell -m}-(-1)^m \, Y_{\ell  m} \right) & \mbox{if $m<0$}
\end{cases}
\end{equation}

The matrix $X^T W X$ is a mixing matrix that couples different spherical harmonic modes together. The effect of the mask $W$ is that a least squares fitting process can then be used on the masked map to determine the dipole from clusters alone. Caution is needed because information is lost when the map is masked. Consequently if too few modes are fitted, power will `leak' from higher order modes into the fit parameters, corrupting the result.

Since the kSZ signal is proportional to the optical depth, in principle the signal to noise ratio of the measurement can be increased by weighting each pixel by the optical depth. We find that the increase in signal to noise is approximately $10\%$. However, when filtering with the MF this weighting scheme increases the contamination from thermal SZ, since the thermal SZ signal is also proportional to the optical depth. We therefore do not weight by the optical depth to reduce the risk of contaminating our result.

In principle the correlated errors on the three dipole $m$ values can be calculated, with covariance matrix

\begin{equation}
\label{eqn:pcov}
N = (X^T W X)^{-1} X^T W C W^T X (X^T W X)^{-1}
\end{equation}

where $C$ is the pixel-pixel noise covariance matrix. Since it is not practical to compute a matrix as large as $C$, we calculate the covariance using simulations. The dominant source of error is CMB and instrument noise, which we estimate by passing 100 CMB and noise realizations through our pipeline, performing the least squares fit on each. There are additional sources of error from our uncertainty in the optical depth, and from the random component of the galaxy cluster peculiar velocities, which are calculated from equation~\ref{eqn:sigv}. We estimate the errors from both of these terms by passing simulations of the kSZ signal through our analysis pipeline, performing the least squares fit on each realization. The scatter in the derived values of $\beta$ then provides an estimate of the noise correlations between the dipole directions, which is then used to calculate the $\chi^2$ significance of the measured dipole in each redshift shell. We find that the errors from the latter two terms are negligible compared to the CMB and noise error, and that the errors on the dipole directions are Gaussian with a narrower distribution in the direction perpendicular to the galactic plane. This is due to clusters lying within the galactic mask being removed from the fit.

As a check on this process, we also estimate the errors from the WMAP maps by re-calculating the dipole fit after rotating our weights map away from the clusters. We rotate the weights map in increments of $10^{\prime}$ in the galactic longitudinal direction. We remove points within $1^{\circ}$ of the cluster center to reduce any residual SZ emission. This leaves us with $2147$ different directions. This process allows us to preserve the angular distribution of the cluster sample in the fit. When either filter is used we find errors that are consistent with those found from the CMB noise realizations. The errors we find by rotating the weights map are less reliable than the those calculated from the CMB and noise realizations, since when we rotate our weights map some regions get rotated into the galactic mask, which we then exclude from the fit. The dipole is therefore calculated from fewer clusters in this scheme, and so the errors we quote are from the CMB and noise realizations.

We find that 100 CMB and noise realizations is sufficient to estimate our errors. Taking a typical error for one of the dipole coefficients in the Wiener filtered maps to be $0.8 \mu$K we find that the standard deviation of the sample variance is $0.09 \mu$K for 100 noise realizations. Even if our errors are $10\%$ too large we would still not detect a significant cluster dipole with our filters.

\subsection{Conversion to Velocity Dipole}
\label{sec:conv_dip}

We create a conversion matrix to calculate the velocity dipole from the temperature dipole,

\begin{equation}
\label{eqn:convert}
\textbf{a}_{\rm{v}} = \textbf{M} \; \textbf{a}_{\rm{T}}
\end{equation}

\noindent where $\textbf{a}_{\rm{v}}$ are the velocity monopole and dipole coefficients, $\textbf{a}_{\rm{T}}$ are the temperature coefficients and $\textbf{M}$ is a $4\times4$ conversion matrix. We calculate $\textbf{M}$ by creating four simulated maps of kSZ signal alone from our cluster sample. In one map all of the clusters are given a monopole velocity. Each of the other three maps has a bulk flow velocity in one of three basis directions. These maps are passed through our analysis pipeline to use the same filtering and velocity fitting processes that are used for the real data. The four fit coefficients from each of the input maps are the elements of each row of $\textbf{M}^{-1}$. The matrix $\textbf{M}$ is close to diagonal in all redshift shells, and the velocity and temperature dipoles are close to alignment, as they should be. To convert from $\mu$K to 1000 km/s we use the following values for the diagonal entries of \textbf{M}: 4.07 (our Wiener filter), 3.41 (KAKE Wiener filter), 0.075 (matched filter), and 0.12 (unbiased matched filter). For the dipole contribution:

\begin{equation}
\textbf{M} = \frac{10^{-12}}{T_{\rm CMB}} \frac{c}{\tau_{\rm eff}} \left( \begin{array}{ccc} 1 & 0 & 0 \\
0 & 1 & 0 \\
0 & 0 & 1 \end{array} \right)
\end{equation}

where $\tau_{\rm eff}$ is the effective optical depth after filtering and we have used the same normalization as KAKE (which makes $\textbf{M}$ a factor of $\sqrt{3/4\pi}$ smaller than for the conventional definition of $a_{1m}$). For our Wiener filter we find $\tau_{\rm eff} = 2.8 \times 10^{-5}$. This value is smaller than the average optical depth of our cluster sample ($\tau = 4.9 \times 10^{-3}$) due to the filtering, as well as the fact that the signal is diluted by beam smoothing and averaging over the aperture used to calculate the cluster dipole. We find a lower effective optical depth than the value used in the KAKE analysis of $\tau_{\rm eff} \sim 10^{-4}$, which could explain why we find velocity limits $\sim 10$ times larger. We are confident that we correctly recover the cluster velocities and show in figure~\ref{fig:amp} that we correctly recover the velocities in simulations.

\subsection{Tests of the Method}
\label{sec:tests}

We have performed tests to check that the results from our filtering and dipole fitting procedures are not contaminated by systematic effects. First we verify that we can correctly recover a known cluster bulk flow from simulated kSZ maps alone, then we consider maps containing kSZ, CMB and instrument noise. In section~\ref{sec:sys} we describe the effects of thermal SZ, unresolved sources, and galactic emission on this process.

\begin{figure}
  \centering
  \includegraphics[width=90mm]{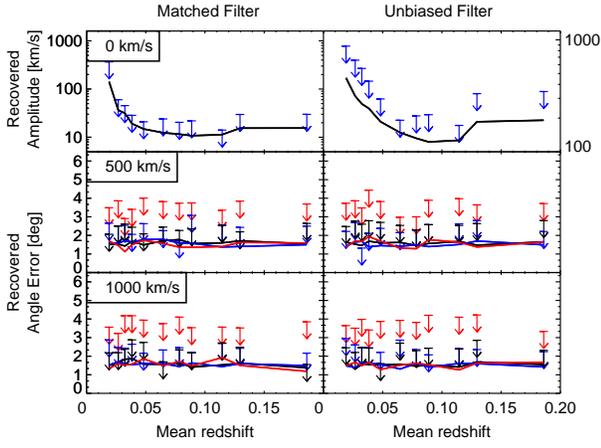}
  \caption{Left: Error in the recovered dipole direction from simulated kSZ maps with a bulk flow velocity of 0 km/s (top), 500 km/s (middle) and 1000 km/s (bottom) in the directions: (latitude, longitude) = ($0^{\circ}$, $0^{\circ}$) black lines, ($0^{\circ}$, $90^{\circ}$) blue lines and ($90^{\circ}$, $0^{\circ}$) red lines, using the MF. Right: Same for the UF. The arrows are the $95\%$ confidence limits.}
  \label{fig:recovered}
\end{figure}

Using simulated maps with a kSZ component only, but that include a cluster bulk flow we have verified that we recover the input bulk flow velocity using our method. We assign all of the clusters in the simulated maps a kSZ signal from both the bulk flow and a random velocity drawn from the expected $\Lambda$CDM distribution in equation~\ref{eqn:sigv}. For each choice of input bulk velocity we generate 25 realizations, filter the maps, and fit for a cluster dipole using the same pipeline that is used for the WMAP data. We repeat the process with three different input bulk flow directions: (galactic latitude, longitude) = ($0^{\circ}$, $0^{\circ}$), ($0^{\circ}$, $90^{\circ}$) and ($90^{\circ}$, $0^{\circ}$) and bulk flow velocities of 0 km/s, 500 km/s and 1000 km/s (7 cases with 25 realizations each). Figure~\ref{fig:recovered} shows the results of the fits. We do not plot the recovered velocity for the cases where the simulated bulk velocity is non-zero because we use our kSZ simulations to calibrate the recovered velocity, and so the mean recovered velocity is exact by design. The scatter in the recovered dipole direction is due to the random component of the cluster line of sight velocity (with variance given by equation~\ref{eqn:sigv}) and uncertainty in the cluster optical depth. The recovered amplitude in the maps with no bulk flow velocity has significantly larger scatter when it is calculated using the UF. This reflects the lower signal to noise of this filter caused by the extra degree of freedom.

\begin{figure}
  \centering
  \includegraphics[width=82mm]{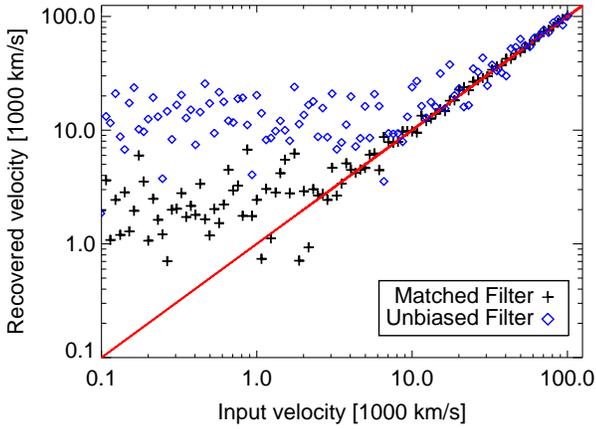}
  \caption{Recovered bulk flow velocity in simulated maps containing CMB, noise and kSZ using the MF (black) and the UF (blue). All clusters are included in the fit. The x-axis is the bulk velocity input into the simulated maps, the y-axis is the recovered velocity. The red line indicates perfect recovery. When the input bulk flow is small, a dipole is not detected and the data points provide an estimate of the scatter in the recovered amplitude.}
  \label{fig:amp}
\end{figure}

\begin{figure}
  \centering
  \includegraphics[width=80mm]{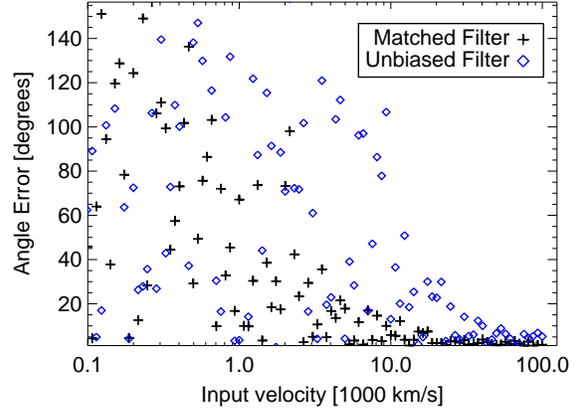}
  \caption{Error in the direction of the bulk flow velocity in simulated maps containing CMB, noise and kSZ with the MF (black) and UF (blue), when all clusters are included in the fit.}
  \label{fig:ang_err}
\end{figure}

\begin{figure}
  \centering
  \includegraphics[width=83mm]{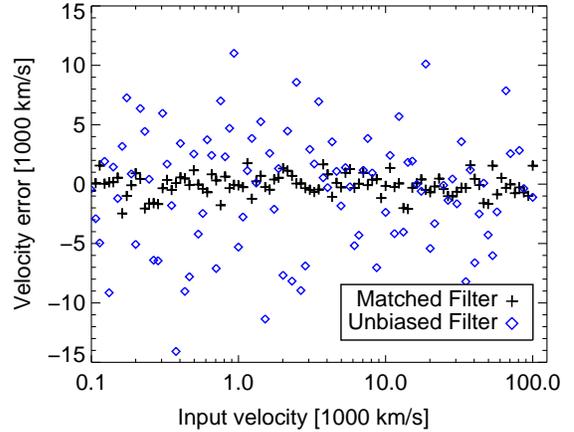}
  \caption{Error in the recovered monopole velocity in simulated maps containing CMB, noise and kSZ with the MF (black) and UF (blue).}
  \label{fig:mono_rec}
\end{figure}

We have repeated the test using simulations that additionally contain CMB and instrument noise. We create 100 realizations for each WMAP channel and add a bulk flow signal to the kSZ component, with a range of velocities logarithmically spaced between 100 and 100,000 km/s in the direction of galactic latitude $-11^{\circ}$ and longitude $103^{\circ}$, which is the bulk flow direction found by KAKE in the redshift shell extending to $z=1$. Figure~\ref{fig:amp} shows the recovered bulk flow amplitude and figure~\ref{fig:ang_err} shows the error in the direction of the recovered dipole when all clusters are included in the fit. The lower signal to noise of the UF is apparent in both of these figures. Figure~\ref{fig:mono_rec} shows the recovered monopole velocity in a different set of 100 simulated WMAP realizations.

\section{Systematic Effects}
\label{sec:sys}

\subsection{Thermal SZ}
\label{sec:sys_tsz}

There are not expected to be intrinsic large scale moments in the tSZ signal, but due to the limited size of our cluster sample the tSZ dipole signal caused by random scatter is not negligible. We have examined the effects of thermal SZ on our results using our own simulations of the tSZ emission from our cluster sample, which are described in section~\ref{sec:sims}, as well as a simulated tSZ map produced using the Planck Sky Model, described in section~\ref{sec:extsim}.

We find that when we fit for the cluster dipole amplitude in our tSZ simulated maps filtered with the MF, using all clusters in our sample, there is a $3\sigma$ bias in the y and z directions, which equates to a dipole amplitude of approximately $4000 \pm 400$ km/s in the direction of $l,b=(88^{\circ}$,$54^{\circ}) \pm 40^{\circ}$, which is $66^{\circ}$ from the KAKE bulk flow direction in the same redshift shell. However, in the PSM tSZ map we find a reduced dipole of 2500 km/s, in the direction of $l,b=(327^{\circ}$,$10^{\circ})$, which is $137^{\circ}$ from the KAKE bulk flow direction. These simulations suggest that contamination to the cluster dipole measurement is at the $1\sigma$ level and so cannot be ignored. As expected, we find that the tSZ monopole in the simulated maps is large and equivalent to a kSZ signal from clusters with velocities of 12000 km/s in our simulations, and 10000 km/s in the PSM maps.

The simulated tSZ maps filtered with the UF have a cluster monopole and dipole amplitude that are more than an order of magnitude lower. In our simulated maps containing only tSZ we find a recovered cluster dipole amplitude of approximately 200 km/s. In the PSM simulated tSZ map we find a recovered cluster dipole amplitude of 70 km/s, significantly less than the measured cluster dipole and noise. The monopole amplitude is 200 km/s in our simulated tSZ map and 100 km/s in the PSM tSZ map.

Although the UF removes the tSZ signal sufficiently for our purposes, it does not remove it to the $\Lambda$CDM limit. The residual tSZ in the filtered maps, and the cause of the non-zero cluster monopole and dipole is source confusion. The emission from clusters with small angular separation overlaps more strongly in the Q band where the beam is large, than in the W band. When the channels are subtracted the tSZ signal does not subtract perfectly in the central cluster pixel and there is residual tSZ in the map. Since this only affects a small number of clusters, it leaves a dipole signal in the map.

\subsection{Radio Point Sources}
\label{sec:sys_radio}

At frequencies below 100 GHz extragalactic radio sources can be a significant contribution to CMB maps. We use simulations of the unresolved extra-galactic radio point source emission produced by~\citet{2010arXiv1001.3659C} and summarized in section~\ref{sec:extsim}, to verify that our bulk flow measurement is not significantly affected by sources. Because the model is based on sources observed by NVSS, the simulated maps retain information about the distribution of sources on the sky and account for the increased radio emission at galaxy cluster positions, caused by clustering of radio galaxies.

The uncertainty on the radio point source monopole and dipole in the maps recovered using the UF are larger than in the maps filtered with the MF. In the maps filtered with the MF we find a non-zero dipole introduced by radio sources with a signal of amplitude $0.2\sigma$ in each of the dipole directions. In the redshift shell encompassing all clusters we find signals of $0.2\sigma$, $0.1\sigma$ and $-0.3\sigma$ in the x, y and z directions previously defined. We find a monopole signal of $-2\sigma$, equivalent to a kSZ monopole velocity of approximately $-2000$ km/s. In the maps filtered with the UF we find a larger signal with an amplitude of $-0.9\sigma$, $1\sigma$ and $1.1\sigma$ in each of the dipole directions. The monopole signal is $-6\sigma$, as expected it is large because the radio sources all add signal to the maps whereas the dipole is caused by sample variance.

Because the UF is designed to remove tSZ by a specific weighting of channels, the radio source contribution is actually amplified by the filter. In the UF the Q band channel is given relatively more weight, so that emission with a tSZ spectrum is canceled when the channels are combined. The radio point source signal is strongest in the Q band and weakest in the W band. The UF therefore has a large contribution from the Q band radio signal that is not offset by the W band signal, where the radio point source signal is much weaker. The resulting map generated using the UF therefore has a greater residual point source signal than the map generated with the MF. The radio point source signal and the tSZ signal could in principle both be suppressed simultaneously by using a filter designed to remove both signals~\citep[e.g.,]{1996MNRAS.281.1297T}. However, the increase in the number of degrees of freedom that this would introduce would further reduce the signal to noise of the measurement.

\subsection{Galactic Emission}
\label{sec:gal}

For our analysis we use the WMAP foreground reduced maps which have been processed by the WMAP team to suppress the galactic signal outside of the WMAP galactic mask~\citep{2010arXiv1001.4744J}. We test the effect that any residual galactic foreground would have on our results by repeating the analysis using maps that have not had the galactic components suppressed. The cluster monopole and dipole that we obtain from these maps is therefore an upper limit on any residual galactic contamination in our results.

\begin{figure}
  \centering
  \includegraphics[width=80mm]{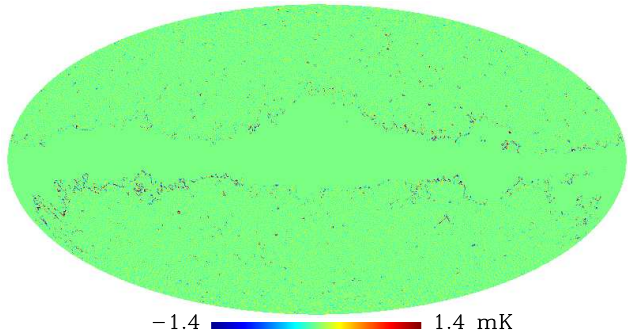}
  \caption{Sum of the PSM simulated maps convolved with the WMAP beams and filtered with the MF. This map shows galactic emission around the edges of the mask that is absent from the WMAP foreground reduced maps due to WMAP's foreground reduction method~\citep{2010arXiv1001.4744J}. The color scale is altered to better show the galactic emission by mimicking the effect of replacing the data by $\sinh^{-1} (\rm{data})$~\citep{2005ApJ...622..759G}.}
  \label{fig:psm_matched}
\end{figure}

\begin{figure}
  \centering
  \includegraphics[width=80mm]{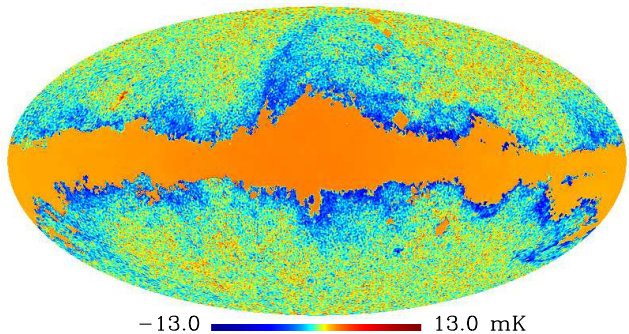}
  \caption{Sum of the PSM simulated maps convolved with the WMAP beams and filtered with the UF. The $\sinh^{-1}$ color scale that was used in figure~\ref{fig:psm_matched} is used here.}
  \label{fig:psm_biasrem}
\end{figure}

Figures~\ref{fig:psm_matched} and~\ref{fig:psm_biasrem} show the sum of the filtered PSM simulated maps. Both maps show residual galactic emission that is not visible in the WMAP maps (figures~\ref{fig:wmapmap_matched_nosz} and~\ref{fig:wmapmap_matched}) due to their foreground reduction procedure. The larger visible foreground signal in the PSM simulated map filtered with the UF is caused by galactic synchrotron. The synchrotron signal is smaller in the map filtered with the MF because the large scale galactic signal is removed at each frequency by the MF, but remains in the maps filtered by the UF. When the filtered maps at each frequency are combined, the signals from the galactic components are not canceled since the filters are designed to only cancel a signal that has a thermal SZ spectrum.

Because the maps containing only the individual contribution of a particular foreground component are available for the PSM simulations, we have also run our analysis pipeline on each component separately to identify the source of any bias. The results when all clusters are included in the dipole fit are shown in Table~\ref{tbl:psm}.

We find no significant bias in the monopole or dipole of the maps filtered by the MF. The combined map filtered by the UF has the largest galactic contribution from synchrotron emission, with a $-5.7 \sigma$ signal in the monopole (labeled $^{\dagger}$ in the table) and a $-4 \sigma$ signal in the dipole direction that points towards the center of the galaxy (labeled $^{\dagger \dagger}$ in the table). The dust and free-free monopole signals largely cancel resulting in a $-6.2 \sigma$ monopole signal. The significant dipole from synchrotron emission is partially canceled by the dust and free-free signals, resulting in a dipole signal that is not significant and would not be considered a detection.

As a further check, we repeat the cluster dipole analysis with the cluster coordinates rotated by one degree away from the true cluster locations. We find that the emission from the galactic components is not significantly changed. Since the emission is not localized at galaxy cluster locations, we expect the galactic signals we find to be suppressed by the WMAP foreground reduction procedure, and so we expect the galactic signals in Table~\ref{tbl:psm} to be an upper limit on residual galactic emission in the WMAP maps.

\begin{deluxetable}{lc|cc|cc|cc|cc|cc}
\rotate
\tabletypesize{\footnotesize}
\tablecolumns{12}
\tablewidth{0pc}
\tablecaption{Monopole and dipole in the PSM simulations.\tablenotemark{*} \label{tbl:psm}}
\tablehead{
	\colhead{Component}				&
	\colhead{Filter}				&
	\multicolumn{2}{c}{Monopole}			&
	\multicolumn{6}{c}{Dipole \tablenotemark{b}}	&
	\colhead{$\sqrt{C_1}$ [$\mu$K]}			&
	\colhead{$95 \%$ Confidence Limit  [$\mu$K]}	\\
	\colhead{}					&
	\colhead{}					&
	\colhead{$a_0$ [$\mu$K]}			&
	\colhead{$\sigma$ \tablenotemark{a}}		&
	\colhead{x [$\mu$K]}				&
	\colhead{$\sigma$ \tablenotemark{a}}		&
	\colhead{y [$\mu$K]}				&
	\colhead{$\sigma$}				&
	\colhead{z [$\mu$K]}				&
	\colhead{$\sigma$}				&
	\colhead{}					&					
	\colhead{}
}
\startdata
Synchrotron  & Matched           & 0.40  & 0.038                    & 0.054 & 0.0020                 & 0.16 & 0.0079 & -0.086 & -0.0048 & 0.19 & 62  \\
Dust         & Matched           & 0.36  & 0.034                    & 0.034 & 0.0013                 & 0.38 & 0.019  & 0.018  & 0.0010  & 0.38 & 62  \\
Free-free    & Matched           & 0.042 & 0.0040                   & -0.46 & -0.017                 & 0.38 & 0.019  & -0.042 & -0.0023 & 0.60 & 62  \\
All galactic & Matched           & 0.80  & 0.077                    & -0.37 & -0.014                 & 0.92 & 0.046  & -0.11  & -0.0061 & 1.00 & 62  \\
Synchrotron  & tSZ Bias Removing & -469  & $-5.7^{\dagger}$         & -717  & $-4.0^{\dagger \dagger}$ & 156  & 1.0    & -120   & -1.0    & 743  & 432 \\
Dust         & tSZ Bias Removing & 58    & 0.70                     & 170   & 0.94                   & -159 & -1.1   & 70     & 0.58    & 244  & 432 \\
Free-free    & tSZ Bias Removing & -101  & -1.2                     & 188   & 1.0                    & 177  & 1.2    & -29    & -0.24   & 260  & 432 \\
All galactic & tSZ Bias Removing & -513  & -6.2                     & -357  & -2.0                   & 174  & 1.2    & -80    & -0.66   & 405  & 432 \\
\enddata
\tablenotetext{*}{In the redshift 0.00-1.00 shell.}
\tablenotetext{a}{The uncertainty estimates are obtained from CMB and noise realizations.}
\tablenotetext{b}{The dipole basis directions are x ($l=0^{\circ}$, $b=0^{\circ}$), y ($l=90^{\circ}$, $b=0^{\circ}$), z ($l=0^{\circ}$, $b=90^{\circ}$).}
\end{deluxetable}

\section{Results}
\label{sec:results}

\subsection{Dipole}

\begin{figure}
  \centering
  \includegraphics[width=85mm]{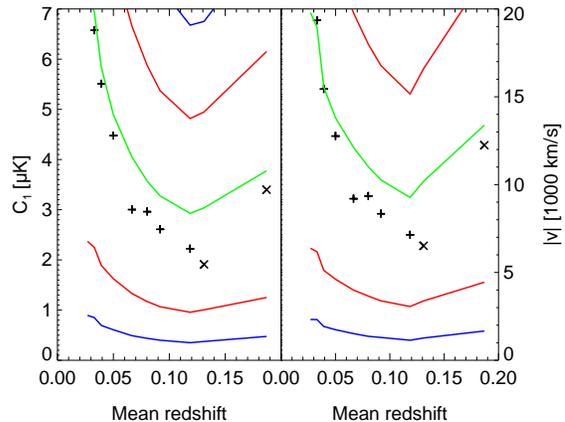}
  \caption{Left: Cluster dipole amplitude in the maps filtered by the KAKE filter. Points with a plus sign have redshift shells with minimum redshift of $0$, those with a cross have shells with minimum redshift of $0.05$ and $0.12$. The green line is the noise bias, the red line is the $95\%$ confidence limit that there is no bulk flow and the blue line is the $99.7\%$ confidence limit. Right: Cluster dipole amplitude in km/s.}
  \label{fig:cl_KY}
\end{figure}

\begin{figure}
  \centering
  \includegraphics[width=85mm]{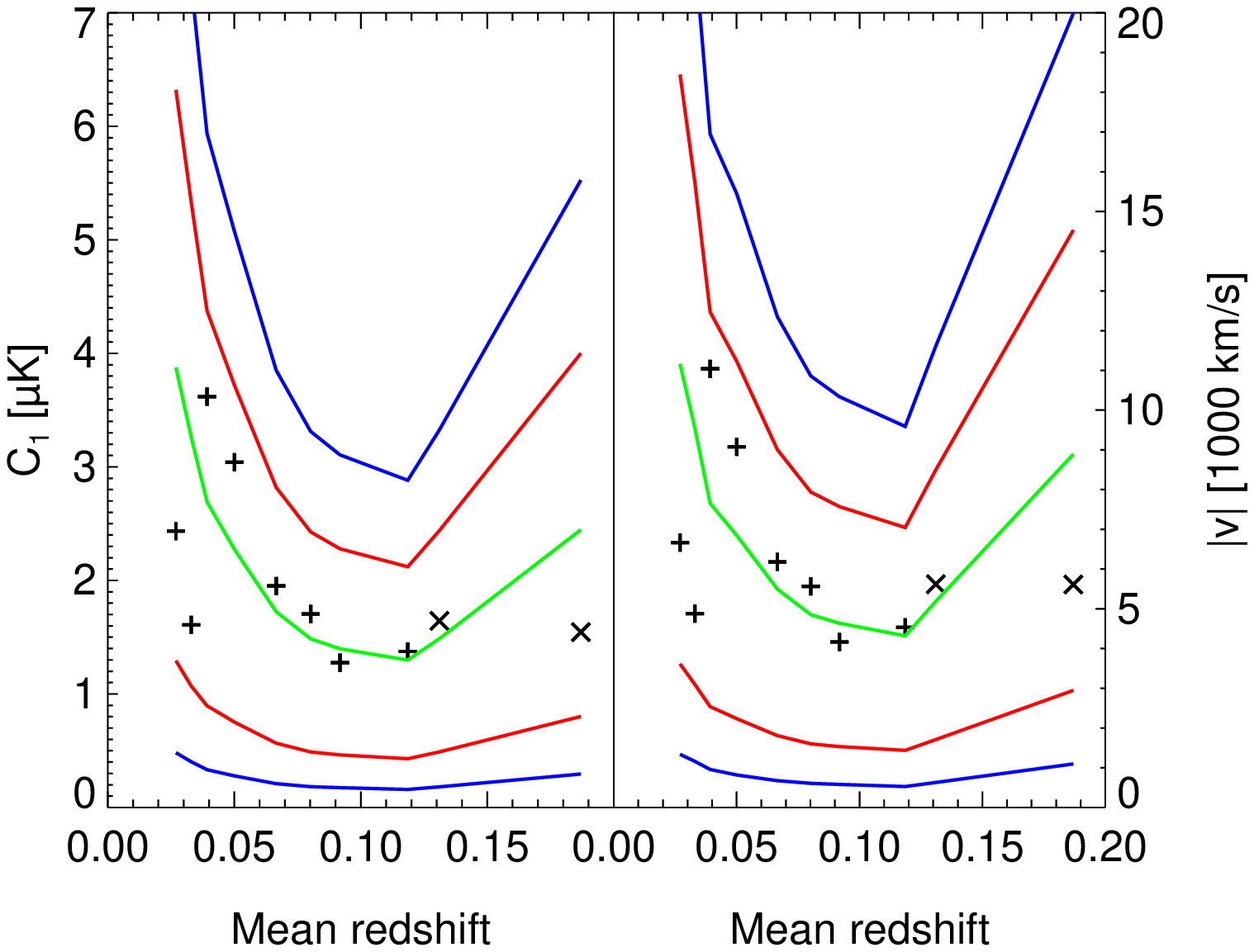}
  \caption{Cluster dipole amplitude in the maps filtered by the WF.}
  \label{fig:cl_WF}
\end{figure}

\begin{figure}
  \centering
  \includegraphics[width=85mm]{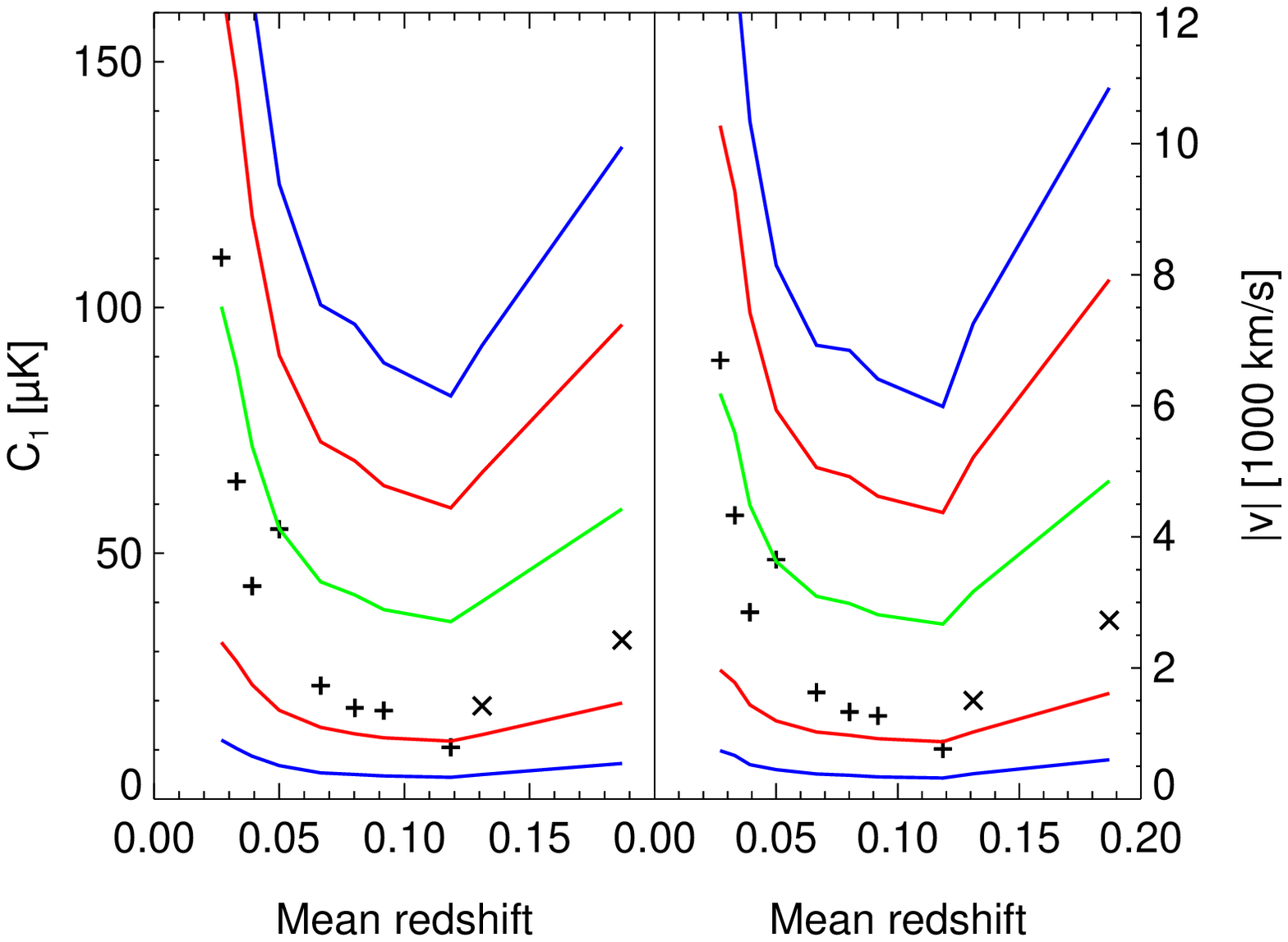}
  \caption{Cluster dipole amplitude in the maps filtered by the MF.}
  \label{fig:cl_nosz}
\end{figure}

\begin{figure}
  \centering
  \includegraphics[width=85mm]{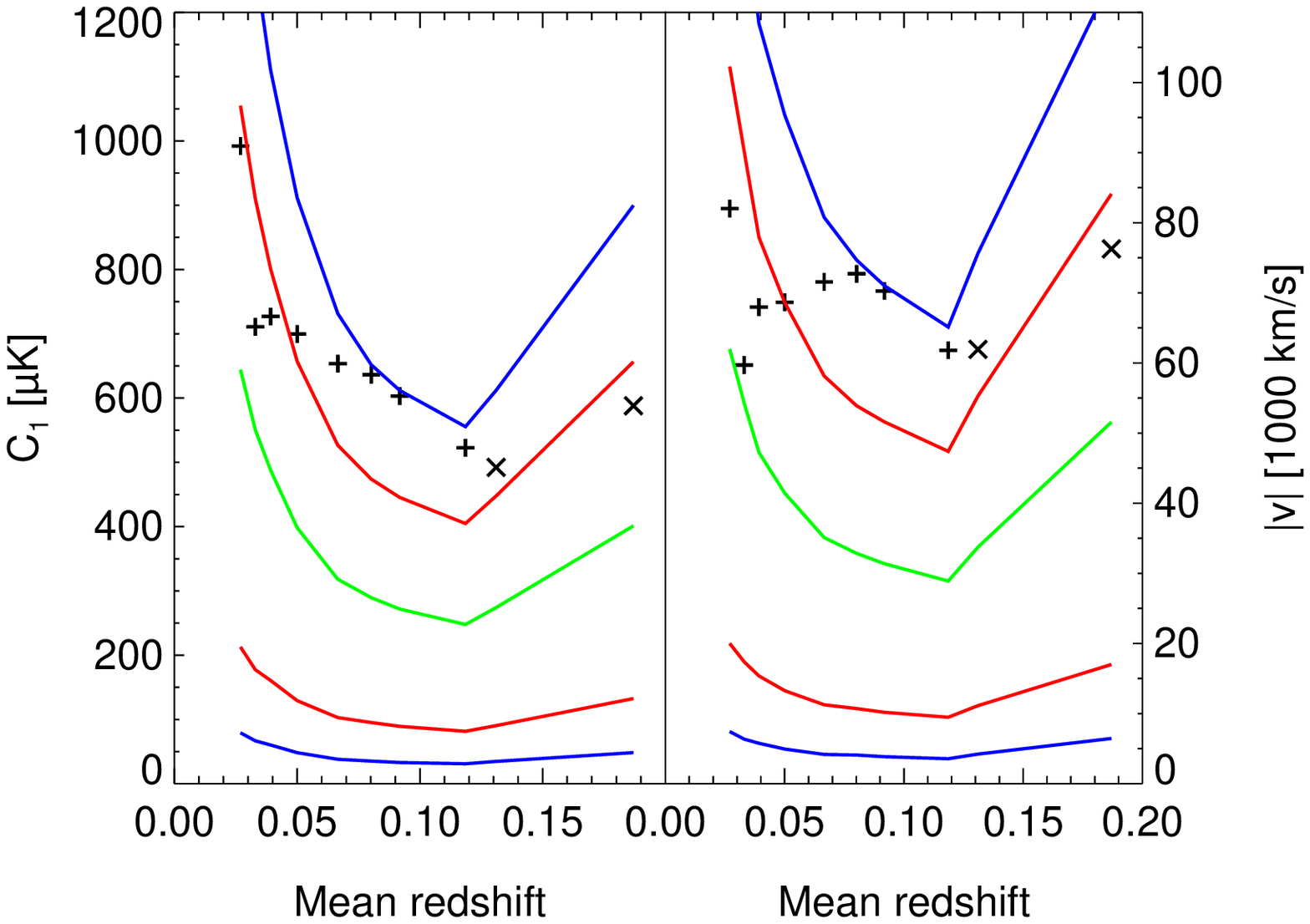}
  \caption{Cluster dipole amplitude in the maps filtered by the UF.}
  \label{fig:cl_mf}
\end{figure}

Table~\ref{tbl:main_KY} and figure~\ref{fig:cl_KY} show our KAKE filter pipeline results, table~\ref{tbl:main_WF} and figure~\ref{fig:cl_WF} show the WF results, table~\ref{tbl:main_nosz} and figure~\ref{fig:cl_nosz} show the MF results and table~\ref{tbl:main_mf} and figure~\ref{fig:cl_mf} show the UF results.  The points in figures~\ref{fig:cl_KY}$-$\ref{fig:cl_mf} are from the WMAP~7 year data. The green line is the noise bias, the red and blue lines are the $95\%$ and $99.7\%$ confidence limits that there is no bulk flow, which are estimated from our realizations containing CMB and instrumental noise. We find smaller errors for our Wiener filter than for the KAKE filter since our filter suppresses more power at $\ell \lesssim 100$ as shown in figure~\ref{fig:filt_comp}. The noise at these multipoles is dominated by the CMB and is correlated between channels~\citep{2009ApJ...707L..42K}. We do not find a significant dipole in any of the redshift shells using any of the filters. In figure~\ref{fig:cl_mf} some of the points are close to the $99.7\%$ confidence limit. Since this filter is much less sensitive than the other filters which have results that are consistent with noise, this result cannot be due to a bulk flow. Instead it is likely that it is at least partly due to radio point source contamination which in section~\ref{sec:sys_radio} we estimated to cause a $\sim 1\sigma$ bias in each of the $a_{1m}$ components in maps filtered by the UF. We have repeated our analysis for the WMAP~5 year data and find no significant cluster dipole. We find that for the MF the CMB and instrument noise contribute approximately equally. For the UF the noise is dominated by the instrument noise, with approximately $90\%$ contribution.

The UF is an order of magnitude less sensitive than the MF. This is due to the channels being combined in a way that is not optimal for maximizing the kSZ signal, but results in cancellation of the tSZ signal. The reduction in sensitivity of the UF is largely due to the limited frequency coverage of the maps we use, and so an experiment with greater frequency coverage, such as Planck, would perform better with this filter.

\begin{figure}
  \centering
  \includegraphics[width=75mm]{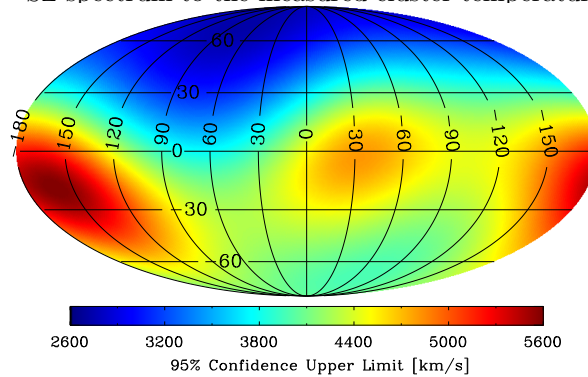}
  \caption{$95\%$ confidence upper limit to the bulk flow in the redshift $0-1$ shell. We find a bias in simulated thermal SZ maps equivalent to a bulk velocity of $\sim \! 2500$ km/s.}
  \label{fig:dipmap}
\end{figure}

Although the dipole amplitude we find is consistent with zero, our limits to the bulk flow velocity are tighter in some directions than in others. Figure~\ref{fig:dipmap} shows the $95\%$ confidence upper limit to the bulk flow over the whole sky using the results from the MF with all clusters included in the fit. Table~\ref{tbl:dipoles} gives the upper limit to the flow in the direction of other well-known dipoles. The upper limit in all directions is above the limit expected by cosmic variance in the $\Lambda$CDM model, and is above the measured low redshift flow.

\setcounter{table}{6}

\begin{deluxetable}{lcccc}
\tabletypesize{\footnotesize}
\tablecolumns{5}
\tablewidth{0pc}
\tablecaption{$95\%$ Confidence Limits on the Dipole.\label{tbl:dipoles}}
\tablehead{
	\colhead{Dipole Name}				&
	\colhead{$l$ [deg]}				&					
	\colhead{$b$ [deg]}				&
	\colhead{Velocity [km/s]}			&
	\colhead{Reference \tablenotemark{1}}
}
\startdata
CMB                 & 263.99  & 48.26   & 3485  & $[1]$   \\
Lauer $\&$ Postman  & 220     & $-$28   & 4625  & $[2]$   \\
Watkins             & 287     & 8       & 4509  & $[3]$   \\
KAKE                & 267     & 34      & 3886  & $[4]$   \\
\enddata
\tablenotetext{1}{References: [1]~\citet{2010arXiv1001.4744J}, [2]~\citet{1994ApJ...425..418L}, [3]~\citet{2009MNRAS.392..743W}, [4]~\citet{2009ApJ...691.1479K}}

\end{deluxetable}

\subsection{Monopole and Higher Moments}

The monopole result from the MF is expected to be strongly contaminated with thermal SZ emission, and we find a result consistent with tSZ contamination: $-6868 \pm 838$ km/s. In the maps filtered with the UF we find a monopole consistent with zero: $4692 \pm 8947$ km/s. We find no moments between multipoles 2 and 5 that are significantly different from zero.

\subsection{Comparison with SuZIE measurements}

The SuZIE II experiment has placed limits on the bulk flow using pointed observations of galaxy clusters~\citep{2003ApJ...592..674B,2004PhDT........35B}. SuZIE II made simultaneous measurements of the SZ effect in three frequency bands, centered at 145 GHz, 221 GHz, and 355 GHz, from the Caltech Submillimeter Observatory, with a $1^{\prime}.5$ beam FWHM at all frequencies. A significant detection of the thermal SZ signal was made in 15 clusters; no significant detection of the kinetic SZ signal was made in any of the clusters observed.

\citet{2003ApJ...592..674B} fit a thermal and kinetic SZ spectrum to the measured cluster temperature at each frequency. No significant cluster dipole was found in the SuZIE cluster sample. We find that the SuZIE bulk flow limits are more sensitive than our results with the MF. We have combined the cluster dipole from the SuZIE data with the dipole limit we find in the WMAP data. When combining the data we weight the dipole amplitude in each of the x, y, and z directions by the inverse noise variance. We find that the $95\%$ confidence limit in the direction of the CMB dipole is increased from the SuZIE result of 1500 km/s to 1800 km/s with the combined data, because our best fit dipole points in a different direction to the SuZIE dipole. With the UF our result is given less than $1\%$ weight, and the combined result is unchanged from the SUZIE dipole measurement. The cluster monopole velocity in the SuZIE data using all of the clusters is $-570 \pm 320$. The result with the UF is not sensitive enough to tighten this constraint.

\section{Conclusions}
\label{sec:conclusions}

We have used the kinetic Sunyaev-Zeldovich effect to look for large scale moments in the galaxy cluster line of sight velocity distribution in the WMAP~7 year data. We use a multi-frequency matched filter that maximizes the cluster signal to noise ratio. We use a sample of 736 clusters derived from the ROSAT X-ray catalogs, and calculate the dipole at the locations of the clusters in the filtered maps.

We find no evidence of a cluster dipole in the WMAP~7 year data in any of the redshift shells we use, consistent with predictions from the $\Lambda$CDM theory. Using kSZ simulations we create a matrix to convert the temperature dipole amplitude to a velocity amplitude. We find a $95\%$ confidence upper limit to the flow of 3485 km/s in the direction of the CMB dipole, and 3886 km/s in the direction of the KAKE claimed flow (galactic longitude $267^{\circ}$, latitude $34^{\circ}$). We have performed our analysis on the WMAP~5 year data used in the KAKE analysis and find no evidence of a cluster bulk flow. We find that results obtained using our Wiener filter have greater sensitivity than the results produced using the KAKE filter, and our matched filter gives results approximately $\sim \! 3$ times more sensitive than those from the KAKE filter. The reason for the difference between our velocity limits and those of KAKE is due to the increased sensitivity of our pipeline and the different factor we use to convert from $\mu$K to km/s, which depends on the kSZ simulations used.

Our analysis differs from that of KAKE in several ways. We use a different filter to suppress the CMB component. We have tried three filters: a Wiener filter, a matched filter and a matched filter that suppresses the tSZ emission. We use a different cluster sample. Our sample contains 736 clusters outside of the galactic mask. The sample described in KAKE has 674 clusters and the sample in~\citet{2010ApJ...712L..81K} contains 985 clusters. For our Wiener filter we do not detect a cluster signal with any aperture we use. The numbers we quote in table~\ref{tbl:main_WF} are from a $15^{\prime}$ aperture which we found gives a higher signal to noise measurement than larger apertures.

Using simulations of the tSZ signal we find that the results with our matched filter are contaminated by thermal SZ, although the contamination is below the WMAP sensitivity. The $\Lambda$CDM model does not predict an intrinsic dipole in the tSZ emission, but due to the relatively small size of our cluster sample, we find a non-zero signal. We estimate the signal to have an amplitude equivalent to a kSZ signal with a bulk flow velocity of $\sim \! 2000-4000$ km/s. The KAKE analysis is performed with large $\sim 30^{\prime}$ apertures around each cluster and so the tSZ signal is diluted, as seen by the small monopole values KAKE observe.

When the maps are filtered with the MF we find a monopole which is consistent with thermal SZ simulations, with a magnitude of $(-91.2 \pm 11.1) \; \mu$K, equivalent to a kSZ signal of $-6868 \pm 838$ km/s. We use a modified multi-frequency matched filter that utilizes the different spectral shapes of the kSZ and tSZ signals to remove the thermal SZ bias. However, the signal to noise of the cluster dipole measurement is reduced by almost an order of magnitude, a consequence of using only three frequency bands in our analysis.

The tSZ bias removing filter also has increased contamination from extra-galactic unresolved radio emission. A filter could be constructed to suppress this signal, however, since this would further reduce the signal to noise of the measurement, we do not further modify our filters to remove it.

The limits we place on the cluster bulk velocity can be decreased by using higher signal to noise measurements of the kinetic SZ signal, as well as by using a larger cluster catalog. We expect that our method can be applied to data from the upcoming Planck experiment. Planck is currently surveying the sky at a resolution of $5^{\prime}$ at 217 GHz. In addition Planck will itself produce a cluster catalog, which can be used for the analysis. Planck's wide frequency coverage from 30-857 GHz, including a channel at the thermal SZ null of 217 GHz, will allow the tSZ signal to be removed with a smaller impact on the kSZ signal to noise ratio. The effect of radio point sources will likely be negligible because the strongest cluster signal is at frequencies where the radio point source signal is small. However, extra-galactic infrared sources with a rising spectrum will probably be an important contaminant. Our filter could be modified using spectral information about the IR source population to suppress emission with this frequency dependence. A paper describing the application of our method to the Planck experiment is in preparation~\citep{2011arXiv1101.1581M}.

\section{Acknowledgements}

We acknowledge very useful discussions with Eiichiro Komatsu. We would like to thank Kris G{\'o}rski and the JPL data analysis group for fostering initial conversations among the authors. We thank Loris Colombo for giving us simulations of the unresolved radio point source emission. SJO would like to acknowledge useful discussions with Neelima Sehgal. EP thanks Stefano Borgani for useful conversations at the beginning of this work and thanks the Aspen Center for Physics for hospitality. SJO and SEC acknowledge support from the US Planck Project, which is funded by the NASA Science Mission Directorate. DSY Mak acknowledges support from the USC Provost's Ph.D Fellowship Program. EP is an ADVANCE fellow (NSF grant AST-0649899). She also acknowledges support from NASA grant NNX07AH59G and JPL-Planck subcontract 1290790. Some of the results in this paper have been derived using the HEALPix~\citep{2005ApJ...622..759G} package. The authors acknowledge the use of the Planck Sky Model, developed by the Component Separation Working Group (WG2) of the Planck Collaboration.

\bibliography{bulk_flow}

\clearpage
\begin{landscape}

\setcounter{table}{2}
\renewcommand{\thefootnote}{\alph{footnote}}

\addtolength{\topmargin}{2in}

\begin{deluxetable}{cccccc cccc c}
\tabletypesize{\footnotesize}
\tablecolumns{11}
\tablewidth{0pc}
\tablecaption{Results from the KAKE filter.\tablenotemark{*} \label{tbl:main_KY}}
\tablehead{
	\colhead{$z_{min}$}									&
	\colhead{$z_{max}$}									&
	\colhead{$\langle z \rangle$ }								&
	\colhead{$z_{median}$}									&
	\colhead{$z_{\sigma}$}									&
	\colhead{$N_{cl}$}									&
	\colhead{Monopole, $a_0$ [$\mu$K] \tablenotemark{a}}					&
	\colhead{$a_{1x}$ [$\mu$K]}								&
	\colhead{$a_{1y}$ [$\mu$K]}								&
	\colhead{$a_{1z}$ [$\mu$K]}								&
	\colhead{Dipole Amplitude, $\sqrt{C_1}$ [$\mu$K] \tablenotemark{b}}
}
\startdata
0.0  & 0.02  & 0.014 & 0.015 & 0.0048 & 28 & -1.4 $\pm$ 3.8 & -12. $\pm$ 10. & -12. $\pm$ 8.7 & -6.6 $\pm$ 8.7 & 18. \\
0.0  & 0.025 & 0.016 & 0.016 & 0.0057 & 36 & -6.2 $\pm$ 3.5 & -8.6 $\pm$ 8.7 & -11. $\pm$ 7.6 & -5.6 $\pm$ 5.5 & 15. \\
0.0  & 0.03  & 0.019 & 0.019 & 0.0075 & 50 & -5.8 $\pm$ 3.1 & -1.8 $\pm$ 6.3 & -4.3 $\pm$ 6.3 & -2.4 $\pm$ 4.4 & 5.2 \\
0.0  & 0.04  & 0.027 & 0.030 & 0.0100 & 95 & -1.6 $\pm$ 2.2 & -3.2 $\pm$ 4.9 & -4.9 $\pm$ 4.3 & -5.6 $\pm$ 3.1 & 8.1 \\
0.0  & 0.05  & 0.033 & 0.035 & 0.012 & 139 & -2.3 $\pm$ 2.1 & -2.1 $\pm$ 4.9 & -6.0 $\pm$ 3.8 & -1.5 $\pm$ 3.0 & 6.6 \\
0.0  & 0.06  & 0.039 & 0.041 & 0.015 & 192 & -3.9 $\pm$ 1.9 & -2.4 $\pm$ 3.8 & -4.9 $\pm$ 3.5 & 0.017 $\pm$ 2.8 & 5.5 \\
0.0  & 0.08  & 0.050 & 0.052 & 0.019 & 294 & -4.6 $\pm$ 1.7 & -1.2 $\pm$ 3.0 & -3.9 $\pm$ 3.0 & -1.9 $\pm$ 2.5 & 4.5 \\
0.0  & 0.12  & 0.067 & 0.066 & 0.029 & 445 & -4.4 $\pm$ 1.5 & -0.82 $\pm$ 2.5 & -2.7 $\pm$ 2.3 & -1.1 $\pm$ 2.2 & 3.0 \\
0.0  & 0.16  & 0.080 & 0.076 & 0.039 & 546 & -3.9 $\pm$ 1.5 & -0.77 $\pm$ 2.2 & -2.7 $\pm$ 2.1 & -0.86 $\pm$ 1.9 & 3.0 \\
0.0  & 0.20  & 0.092 & 0.083 & 0.049 & 619 & -3.6 $\pm$ 1.4 & 0.024 $\pm$ 1.9 & -2.6 $\pm$ 2.0 & -0.31 $\pm$ 1.7 & 2.6 \\
0.0  & 1.0   & 0.12  & 0.097 & 0.079 & 736 & -3.6 $\pm$ 1.2 & 0.55 $\pm$ 1.7 & -2.1 $\pm$ 1.8 & -0.26 $\pm$ 1.6 & 2.2 \\
0.05 & 0.30  & 0.13  & 0.12  & 0.064 & 578 & -3.9 $\pm$ 1.3 & 1.2 $\pm$ 1.7 & -1.4 $\pm$ 1.9 & -0.098 $\pm$ 1.7 & 1.9 \\
0.12 & 0.30  & 0.19  & 0.18  & 0.048 & 271 & -2.1 $\pm$ 1.5 & 2.9 $\pm$ 2.2 & -1.8 $\pm$ 2.4 & 0.56 $\pm$ 2.0 & 3.4 \\
\enddata
\end{deluxetable}

\setcounter{table}{3}

\begin{deluxetable}{cccccc cccc c}
\tabletypesize{\footnotesize}
\tablecolumns{11}
\tablewidth{0pc}
\tablecaption{Results from our Wiener filter.\tablenotemark{*} \label{tbl:main_WF}}
\tablehead{
	\colhead{$z_{min}$}									&
	\colhead{$z_{max}$}									&
	\colhead{$\langle z \rangle$ }								&
	\colhead{$z_{median}$}									&
	\colhead{$z_{\sigma}$}									&
	\colhead{$N_{cl}$}									&
	\colhead{Monopole, $a_0$ [$\mu$K] \tablenotemark{a}}					&
	\colhead{$a_{1x}$ [$\mu$K]}								&
	\colhead{$a_{1y}$ [$\mu$K]}								&
	\colhead{$a_{1z}$ [$\mu$K]}								&
	\colhead{Dipole Amplitude, $\sqrt{C_1}$ [$\mu$K] \tablenotemark{b}}
}
\startdata
0.0  & 0.02  & 0.014 & 0.015 & 0.0048 & 28 & 0.40 $\pm$ 2.5 & -4.8 $\pm$ 5.3 & 0.73 $\pm$ 5.6 & -0.97 $\pm$ 5.0 & 5.0 \\
0.0  & 0.025 & 0.016 & 0.016 & 0.0057 & 36 & -1.9 $\pm$ 2.2 & -3.8 $\pm$ 4.7 & -2.1 $\pm$ 5.0 & -3.2 $\pm$ 3.7 & 5.4 \\
0.0  & 0.03  & 0.019 & 0.019 & 0.0075 & 50 & -3.0 $\pm$ 1.9 & 0.066 $\pm$ 3.9 & 0.15 $\pm$ 3.9 & -1.4 $\pm$ 2.8 & 1.4 \\
0.0  & 0.04  & 0.027 & 0.030 & 0.0100 & 95 & -2.5 $\pm$ 1.3 & -0.58 $\pm$ 2.3 & -1.7 $\pm$ 2.5 & -1.6 $\pm$ 1.9 & 2.4 \\
0.0  & 0.05  & 0.033 & 0.035 & 0.012 & 139 & -2.2 $\pm$ 1.0 & -0.92 $\pm$ 2.1 & -1.3 $\pm$ 2.0 & 0.096 $\pm$ 1.5 & 1.6 \\
0.0  & 0.06  & 0.039 & 0.041 & 0.015 & 192 & -3.4 $\pm$ 0.91 & -2.3 $\pm$ 1.6 & -1.8 $\pm$ 1.7 & 2.1 $\pm$ 1.3 & 3.6 \\
0.0  & 0.08  & 0.050 & 0.052 & 0.019 & 294 & -3.8 $\pm$ 0.79 & -2.3 $\pm$ 1.4 & -1.9 $\pm$ 1.5 & -0.59 $\pm$ 1.0 & 3.0 \\
0.0  & 0.12  & 0.067 & 0.066 & 0.029 & 445 & -3.5 $\pm$ 0.62 & -1.6 $\pm$ 1.1 & -1.1 $\pm$ 1.1 & -0.29 $\pm$ 0.78 & 2.0 \\
0.0  & 0.16  & 0.080 & 0.076 & 0.039 & 546 & -3.7 $\pm$ 0.55 & -1.3 $\pm$ 0.91 & -1.1 $\pm$ 0.96 & -0.41 $\pm$ 0.68 & 1.7 \\
0.0  & 0.20  & 0.092 & 0.083 & 0.049 & 619 & -3.7 $\pm$ 0.51 & -0.91 $\pm$ 0.85 & -0.66 $\pm$ 0.91 & -0.59 $\pm$ 0.63 & 1.3 \\
0.0  & 1.0   & 0.12  & 0.097 & 0.079 & 736 & -3.7 $\pm$ 0.45 & -0.87 $\pm$ 0.80 & -0.95 $\pm$ 0.82 & -0.47 $\pm$ 0.61 & 1.4 \\
0.05 & 0.30  & 0.13  & 0.12  & 0.064 & 578 & -4.1 $\pm$ 0.50 & -1.1 $\pm$ 0.96 & -0.93 $\pm$ 0.91 & -0.74 $\pm$ 0.68 & 1.6 \\
0.12 & 0.30  & 0.19  & 0.18  & 0.048 & 271 & -4.0 $\pm$ 0.66 & -0.19 $\pm$ 1.6 & -1.2 $\pm$ 1.4 & -0.99 $\pm$ 1.2 & 1.5 \\
\enddata
\tablenotetext{*}{The monopole and dipole are calculated within $15^{\prime}$ of the cluster centers. In tables~\ref{tbl:main_nosz} and~\ref{tbl:main_mf} the monopole and dipole are calculated in the central pixel of the clusters.}
\end{deluxetable}

\setcounter{table}{4}

\begin{deluxetable}{cccccc cccc c}
\tabletypesize{\footnotesize}
\tablecolumns{11}
\tablewidth{0pc}
\tablecaption{Results from the matched filter.\label{tbl:main_nosz}}
\tablehead{
	\colhead{$z_{min}$}									&
	\colhead{$z_{max}$}									&
	\colhead{$\langle z \rangle$ }								&
	\colhead{$z_{median}$}									&
	\colhead{$z_{\sigma}$}									&
	\colhead{$N_{cl}$}									&
	\colhead{Monopole, $a_0$ [$\mu$K] \tablenotemark{a}}					&
	\colhead{$a_{1x}$ [$\mu$K]}								&
	\colhead{$a_{1y}$ [$\mu$K]}								&
	\colhead{$a_{1z}$ [$\mu$K]}								&
	\colhead{Dipole Amplitude, $\sqrt{C_1}$ [$\mu$K] \tablenotemark{b}}
}
\startdata
0.0  & 0.02 & 0.014 & 0.015 & 0.0048 & 28 & 59 $\pm$ 65 & -39 $\pm$ 117 & -10 $\pm$ 140 & -56 $\pm$ 101 & 69 \\
0.0  & 0.025 & 0.016 & 0.016 & 0.0057 & 36 & 55 $\pm$ 57 & -111 $\pm$ 110 & -79 $\pm$ 126 & -68 $\pm$ 70 & 152 \\
0.0  & 0.03 & 0.019 & 0.019 & 0.0075 & 50 & -15 $\pm$ 44 & -82 $\pm$ 97 & -110 $\pm$ 101 & -1.8 $\pm$ 60 & 137 \\
0.0  & 0.04 & 0.027 & 0.030 & 0.0100 & 95 & -22 $\pm$ 33 & -20 $\pm$ 60 & -107 $\pm$ 70 & -16 $\pm$ 40 & 110 \\
0.0  & 0.05 & 0.033 & 0.035 & 0.012 & 139 & -41 $\pm$ 27 & -45 $\pm$ 58 & -44 $\pm$ 55 & 13 $\pm$ 36 & 65 \\
0.0  & 0.06 & 0.039 & 0.041 & 0.015 & 192 & -62 $\pm$ 24 & -16 $\pm$ 49 & -23 $\pm$ 43 & 33 $\pm$ 29 & 43 \\
0.0  & 0.08 & 0.050 & 0.052 & 0.019 & 294 & -63 $\pm$ 18 & -39 $\pm$ 37 & -38 $\pm$ 32 & -7.1 $\pm$ 24 & 55 \\
0.0  & 0.12  & 0.067 & 0.066 & 0.029 & 445 & -73 $\pm$ 13 & -22 $\pm$ 28 & -6.1 $\pm$ 28 & -0.059 $\pm$ 20 & 23 \\
0.0  & 0.16  & 0.080 & 0.076 & 0.039 & 546 & -81 $\pm$ 12 & -18 $\pm$ 28 & 1.8 $\pm$ 25 & -1.4 $\pm$ 18 & 19 \\
0.0  & 0.20  & 0.092 & 0.083 & 0.049 & 619 & -85 $\pm$ 12 & -14 $\pm$ 26 & 5.7 $\pm$ 23 & -9.3 $\pm$ 17 & 18 \\
0.0  & 1.0   & 0.12  & 0.097 & 0.079 & 736 & -91 $\pm$ 11 & -5.7 $\pm$ 24 & -1.3 $\pm$ 21 & -8.7 $\pm$ 17 & 11 \\
0.05 & 0.30  & 0.13  & 0.12  & 0.064 & 578 & -99 $\pm$ 12 & -2.2 $\pm$ 26 & 9.5 $\pm$ 24 & -16 $\pm$ 19 & 19 \\
0.12 & 0.30  & 0.19  & 0.18  & 0.048 & 271 & -112 $\pm$ 17 & 11 $\pm$ 37 & -0.13 $\pm$ 37 & -30 $\pm$ 27 & 32 \\
\enddata
\end{deluxetable}

\setcounter{table}{5}

\begin{deluxetable}{cccccc cccc c}
\tabletypesize{\footnotesize}
\tablecolumns{11}
\tablewidth{0pc}
\tablecaption{Results from the tSZ bias removing filter.\label{tbl:main_mf}}
\tablehead{
	\colhead{$z_{min}$}									&
	\colhead{$z_{max}$}									&
	\colhead{$\langle z \rangle$ }								&
	\colhead{$z_{median}$}									&
	\colhead{$z_{\sigma}$}									&
	\colhead{$N_{cl}$}									&
	\colhead{Monopole, $a_0$ [$\mu$K]}							&
	\colhead{$a_{1x}$ [$\mu$K]}								&
	\colhead{$a_{1y}$ [$\mu$K]}								&
	\colhead{$a_{1z}$ [$\mu$K]}								&
	\colhead{Dipole Amplitude, $\sqrt{C_1}$ [$\mu$K] \tablenotemark{b}}
}
\startdata
0.0  & 0.02  & 0.014 & 0.015 & 0.0048 & 28 & 118 $\pm$ 425 & -242 $\pm$ 869 & -1045 $\pm$ 866 & -29 $\pm$ 746 & 1073 \\
0.0  & 0.025 & 0.016 & 0.016 & 0.0057 & 36 & -156 $\pm$ 326 & 269 $\pm$ 756 & -1268 $\pm$ 862 & -449 $\pm$ 537 & 1373 \\
0.0  & 0.03  & 0.019 & 0.019 & 0.0075 & 50 & 73 $\pm$ 267 & 1256 $\pm$ 598 & -29 $\pm$ 610 & -93 $\pm$ 420 & 1259 \\
0.0  & 0.04  & 0.027 & 0.030 & 0.0100 & 95 & -223 $\pm$ 193 & 938 $\pm$ 410 & -281 $\pm$ 384 & 159 $\pm$ 314 & 992 \\
0.0  & 0.05  & 0.033 & 0.035 & 0.012 & 139 & -164 $\pm$ 169 & 616 $\pm$ 394 & -310 $\pm$ 298 & 170 $\pm$ 241 & 710 \\
0.0  & 0.06  & 0.039 & 0.041 & 0.015 & 192 & -22 $\pm$ 140 & 694 $\pm$ 344 & -200 $\pm$ 263 & -74 $\pm$ 224 & 726 \\
0.0  & 0.08  & 0.050 & 0.052 & 0.019 & 294 & -60 $\pm$ 122 & 660 $\pm$ 280 & -230 $\pm$ 228 & 15 $\pm$ 166 & 699 \\
0.0  & 0.12  & 0.067 & 0.066 & 0.029 & 445 & -38 $\pm$ 102 & 607 $\pm$ 225 & -200 $\pm$ 180 & -131 $\pm$ 133 & 653 \\
0.0  & 0.16  & 0.080 & 0.076 & 0.039 & 546 & -56 $\pm$ 95 & 591 $\pm$ 199 & -147 $\pm$ 159 & -182 $\pm$ 135 & 636 \\
0.0  & 0.20  & 0.092 & 0.083 & 0.049 & 619 & -14 $\pm$ 89 & 549 $\pm$ 183 & -123 $\pm$ 154 & -214 $\pm$ 128 & 603 \\
0.0  & 1.0   & 0.12  & 0.097 & 0.079 & 736 & 40 $\pm$ 77 & 486 $\pm$ 165 & -68 $\pm$ 140 & -178 $\pm$ 119 & 522 \\
0.05 & 0.30  & 0.13  & 0.12  & 0.064 & 578 & 98 $\pm$ 92 & 367 $\pm$ 185 & 66 $\pm$ 150 & -320 $\pm$ 134 & 492 \\
0.12 & 0.30  & 0.19  & 0.18  & 0.048 & 271 & 189 $\pm$ 129 & 230 $\pm$ 272 & 343 $\pm$ 230 & -418 $\pm$ 184 & 588 \\
\enddata
\tablenotetext{a}{The monopole is contaminated by foregrounds with levels consistent with simulations.}
\tablenotetext{b}{The errors have a one sided distribution and so we use a $95\%$ confidence limit and do not state error bars.}
\tablenotetext{c}{The modified KAKE filter is not normalized (see discussion in text).}
\end{deluxetable}


\clearpage
\end{landscape}

\renewcommand{\theequation}{A-\arabic{equation}}
\setcounter{equation}{0}
\setcounter{table}{7}

\appendix
\section{Expected Cluster Velocity Dipole}
\label{sec:dipolederiv}

We follow the derivation for the density distribution in~\citet{1973ApJ...185..413P}. The line of sight velocity field can be expanded as

\begin{equation}
a_{\ell m} = \int dr r^2 d \Omega_r \phi(r) Y^{\ast}_{\ell m}(\hat{\textbf{r}}) \textbf{v}(\textbf{r}) \cdot \hat{\textbf{r}}
\end{equation}

where $r$ is the comoving radial distance, $\phi(r)$ is the comoving number density of objects in the sample, which we assume to be isotropic, $\textbf{v}$ is the object peculiar velocity and $Y_{\ell m}(\hat{\textbf{r}})$ are the spherical harmonics. The peculiar velocity at a given wavenumber is related to the over-density by

\begin{equation}
\label{dvrel}
\textbf{v}(\textbf{k}) = i f H_0 \delta(\textbf{k}) \frac{\hat{\textbf{k}}}{k}
\end{equation}

where $f = (a/D) \; dD/da \approx \Omega_m^{0.55}$, $\Omega_m$ is the matter density parameter, $a$ is the scale factor, $D$ is the growth function and $H_0$ is the Hubble constant. The expansion in terms of spherical harmonics is given by

\begin{equation}
\begin{split}
\langle |a_{\ell m}|^2 \rangle &= f^2 H_0^2 \int \frac{d^3k}{(2\pi)^3} \frac{P(k)}{k^2} \left| \int d^3r \phi(r) Y_{\ell m} (\hat{\textbf{r}}) \; \hat{\textbf{k}} \cdot \hat{\textbf{r}} \; e^{-i \textbf{k} \cdot \textbf{r}} \right| ^2 \\
&= 16\pi^2 f^2 H_0^2 \int \frac{d^3k}{(2\pi)^3} \frac{P(k)}{k^2} \left| \int d^3r  \phi(r) Y_{\ell m} (\hat{\textbf{r}}) \; \hat{\textbf{k}} \cdot \hat{\textbf{r}} \; \sum_{{\ell}^{\prime}m^{\prime}} (-i)^{{\ell}^{\prime}} j_{{\ell}^{\prime}}(kr) Y_{{\ell}^{\prime}m^{\prime}} (\hat{\textbf{k}}) Y_{{\ell}^{\prime}m^{\prime}}^{\ast} (\hat{\textbf{r}}) \right| ^2
\end{split}
\end{equation}

\noindent where $P(k)$ is the matter power spectrum and $j_{\ell} (k r)$ are the spherical Bessel functions. Then

\begin{equation}
C_{\ell} = \langle | a_{\ell m} |^2 \rangle = \frac{2}{\pi} f^2 H_0^2 \int dk P(k) \left( \int dr r^2 \phi(r) g_{\ell}(kr) \right)^2
\end{equation}

\noindent where

\begin{equation}
g_{\ell}(kr) = \frac{1}{2\ell+1} \left[ \ell j_{\ell}(kr) - (\ell+1) j_{\ell+1}(kr) \right]
\end{equation}

The selection function, $\phi(r)$, is estimated by calculating the number density of clusters in our sample in radial bins, which we then interpolate to give a smooth function and normalize.

\section{Comparison with KAKE Results}
\label{sec:kcomp}

\subsection{Filter Pipeline Comparison}

\begin{figure}[H]
  \centering
  \includegraphics[width=120mm]{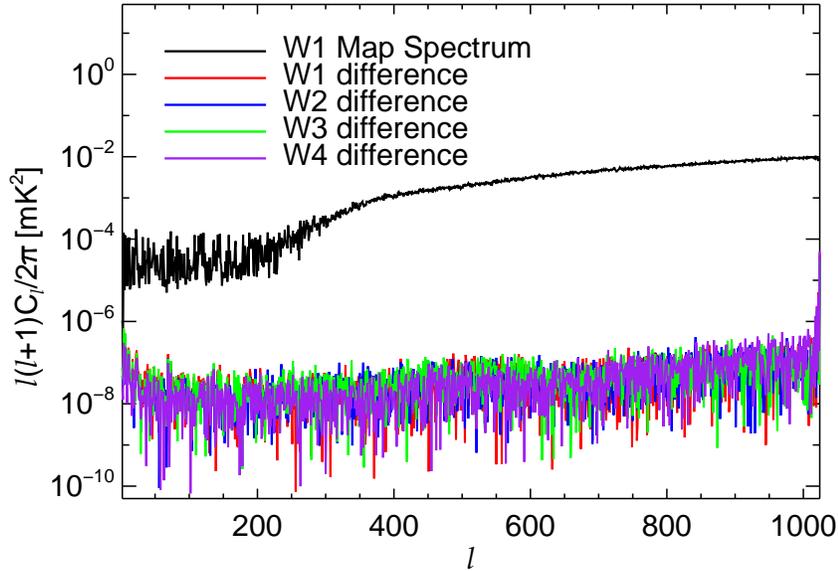}
  \caption{Spectrum of the W1 channel map filtered by the KAKE filter (black) and the difference between the filtered W band map spectra from our KAKE filter pipeline and the spectra of the publicly available maps used in the~\citet{2011ApJ...732....1K} analysis.}
  \label{fig:spec_comp}
\end{figure}

\begin{figure}[H]
  \centering
  \includegraphics[width=120mm]{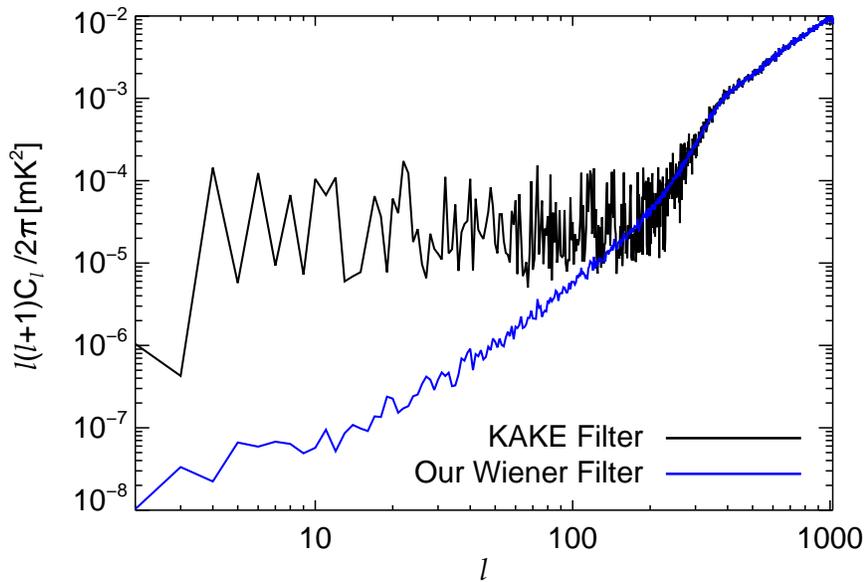}
  \caption{Filtered W1 channel map spectrum from the KAKE pipeline and our pipeline.}
  \label{fig:filt_comp}
\end{figure}

We have verified that we can reproduce the KAKE pipeline by comparing our filtered maps with the publicly available maps used for the~\citet{2011ApJ...732....1K} analysis. We find that the mean difference between the W band maps passed through our pipeline and the filtered maps from the KAKE pipeline is $-1.5 \times 10^{-10} m$K indicating that our pipeline can reproduce their results. As a further check we calculate the spectrum of the filtered W1 channel map, and the absolute value of the difference between our filtered map spectra and the spectra from the KAKE maps in figure~\ref{fig:spec_comp}. In figure~\ref{fig:filt_comp} we show the spectra of the WMAP maps passed through the two pipelines. The signal to noise is low at $\ell < 100$, which is why both filters suppress these multipoles. The KAKE filter is larger at $\ell < 100$ because the filter does not remove the fluctuations in the map that are due to cosmic variance. This leaves noise in the maps that is correlated between channels~\citep{2009ApJ...707L..42K}. By further suppressing the map at $\ell < 100$ our filter reduces the noise in our dipole measurement by a factor of $\sim 2$ while not significantly affecting the signal, as can be seen in figure~\ref{fig:comp_ksz}.

\begin{figure}[H]
  \centering
  \includegraphics[width=120mm]{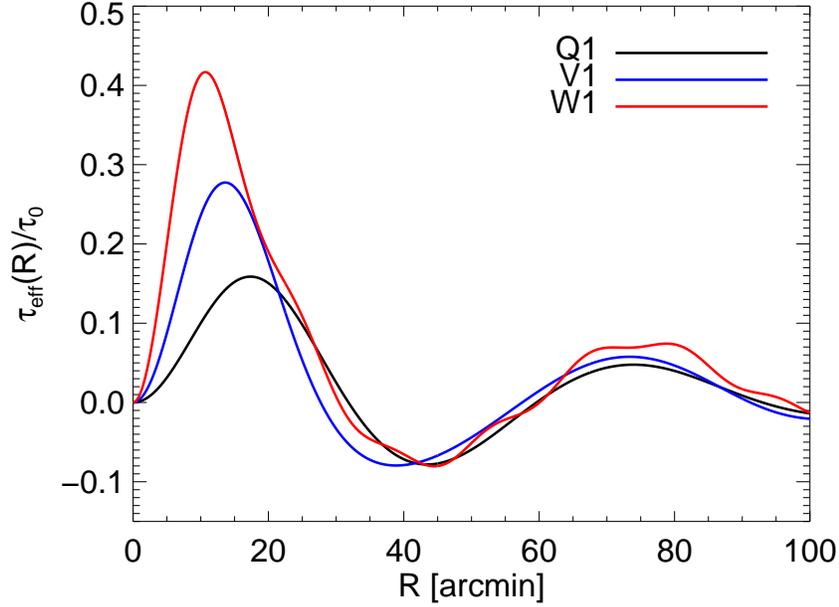}
  \caption{Integrated signal from a beam smoothed and filtered cluster in our simulated maps.}
  \label{fig:filt_prof}
\end{figure}

Since we model our clusters as point sources we can calculate the profile of the cluster signal in the filtered maps. In figure~\ref{fig:filt_prof} we show the integrated signal out to a given radius in the Q1, V1, and W1 channels after filtering with the Wiener filter. The signal is normalized so that if no filtering were applied the result would approach unity at large radius. The fact that the signal peaks at $10-15^{\prime}$ justifies our choice of a $15^{\prime}$ aperture within which to calculate the monopole and dipole.

\subsection{Cluster Dipole Comparison}

In this section we remove clusters with bolometric luminosity $L_X < 2 \times 10^{44} \rm erg/s$ from our sample and use the WMAP Kp0 mask instead of the 7 year extended temperature analysis mask to better compare with the KAKE results. As before we remove the monopole and dipole of the map outside of the mask before fitting for the cluster monopole and dipole. We find that the sample of clusters satisfying the luminosity cut and the redshift cut of $z \le 0.16, 0.2, 0.25$ is almost identical to the sample used in the KAKE analysis. There are some differences in the REFLEX, BCS and BCSe catalogs for clusters with luminosities close to $L_X = 2 \times 10^{44} \rm erg/s$, and some differences for clusters in the CIZA catalog that are near to the WMAP galactic mask boundary. We show the differences we find in table~\ref{tbl:clusdif}.

\begin{table}[H]
  \begin{tabular}{ | l | c | c | c | }
    \hline
    \multirow{2}{*}{$z_{max}$} & Number of clusters & Number of clusters KAKE & Number of clusters we \\
    & with $L_X > 2 \times 10^{44} \rm erg/s$ & include but we don't & include but KAKE do not \\
    \hline
0.16      & 131                & 13                              & 11 \\
0.20      & 205                & 14                              & 18 \\
0.25      & 269                & 14                              & 20 \\
    \hline
  \end{tabular}
  \caption{Number of clusters satisfying the luminosity and redshift cut in our sample and the KAKE sample.}
  \label{tbl:clusdif}
\end{table}

To check that these differences do not change the results by a large amount we have calculated the cluster dipole using the KAKE filter with both the KAKE cluster sample and our own sample. We find the results shown in table~\ref{tbl:wfres} when averaging over the Q, V and W bands.

\begin{table}[H]
  \begin{tabular}{ | lll | cc | cc | cc | cc | }
    \hline
             &           &           & \multicolumn{8}{|c|}{Temperature [$\mu$K]} \\
    \hline
    Cluster sample & $z_{min}$ & $z_{max}$ & m     & $\sigma_m$ & x    & $\sigma_x$ & y     & $\sigma_y$ & z     & $\sigma_z$  \\
    \hline
Ours & 0.0 & 0.16 & -0.74 & 2.05 & 1.44 & 2.87 & -7.04 & 2.98 & 0.04  & 2.68  \\
Ours & 0.0 & 0.20 & 0.02  & 1.48 & 3.04 & 2.21 & -6.23 & 2.52 & 1.93  & 1.97  \\
Ours & 0.0 & 0.25 & 0.04  & 1.21 & 1.56 & 1.92 & -4.17 & 2.14 & 2.08  & 1.72  \\
KAKE & 0.0 & 0.16 & -1.07 & 2.07 & 1.51 & 2.96 & -7.00 & 2.86 & 0.48  & 2.72  \\
KAKE & 0.0 & 0.20 & -0.22 & 1.46 & 3.06 & 2.22 & -6.20	& 2.45 & 2.08  & 2.02 \\
KAKE & 0.0 & 0.25 & -0.20 & 1.22 & 1.68 & 1.99 & -4.25 & 2.12 & 2.19  & 1.76  \\
    \hline
  \end{tabular}
  \caption{Comparison of the cluster monopole and dipole in our cluster sample with the KAKE sample.}
  \label{tbl:wfres}
\end{table}

All of the results have less than $3\sigma$ significance with some of the $a_{1y}$ components having $\sim 2.5\sigma$ significance. The results averaged over the W band channels only (but with the same filter and cluster samples used to produce table~\ref{tbl:wfres}) are shown in table~\ref{tbl:wfwband}.

\begin{table}[H]
  \begin{tabular}{ | lll | cc | cc | cc | cc | }
    \hline
             &           &           & \multicolumn{8}{|c|}{Temperature [$\mu$K]} \\
    \hline
    Cluster sample & $z_{min}$ & $z_{max}$ & m     & $\sigma_m$ & x    & $\sigma_x$ & y     & $\sigma_y$ & z     & $\sigma_z$  \\
    \hline
Ours & 0.0 & 0.16 & -1.64 & 2.11 & 1.11 & 3.12 & -8.63 & 3.24 & 0.01 & 2.84 \\
Ours & 0.0 & 0.20 & -0.41 & 1.59 & 2.71 & 2.42 & -7.73 & 2.67 & 2.08 & 2.04 \\
Ours & 0.0 & 0.25 & -0.49 & 1.30 & 1.03 & 2.06 & -5.53 & 2.28 & 2.10 & 1.81 \\
KAKE & 0.0 & 0.16 & -1.64 & 2.13 & 0.96 & 3.20 & -8.15 & 3.11 & 0.54 & 2.88 \\
KAKE & 0.0 & 0.20 & -0.48 & 1.55 & 2.60 & 2.43 & -7.45 & 2.60 & 2.28 & 2.12 \\
KAKE & 0.0 & 0.25 & -0.60 & 1.30 & 1.06 & 2.13 & -5.43 & 2.26 & 2.25 & 1.87 \\
    \hline
  \end{tabular}
  \caption{Same as table~\ref{tbl:wfres} for the W band channels only.}
  \label{tbl:wfwband}
\end{table}

These results have greater significance than the combined channel results. We still find nothing with more than $3\sigma$ significance, although the $z=0.0-0.20$ shell $a_{1y}$ value has $2.9\sigma$ significance. The result from~\citet{2011ApJ...732....1K} is shown in table~\ref{tbl:kakeres}.

\begin{table}[H]
  \begin{tabular}{ | ll | c | c | c | c | }
    \hline
          &           & \multicolumn{4}{|c|}{Temperature [$\mu$K]} \\
    \hline
    $z_{min}$ & $z_{max}$ & m            & x          & y          & z \\
    \hline
0.00      & 0.16      & -1.47  & 1.20  & -8.26 & 0.38  \\
0.00      & 0.16      & -0.32  & 2.83  & -7.58 & 2.13  \\
0.00      & 0.16      & -0.44  & 1.30  & -5.57 & 2.10  \\
    \hline
  \end{tabular}
  \caption{Result from~\citep{2011ApJ...732....1K}.}
  \label{tbl:kakeres}
\end{table}

which is similar to our result above. These results are more significant than those obtained using the full cluster sample. This cannot be fully explained by the higher luminosity clusters having larger optical depths. In table~\ref{tbl:avgtau} we show the average optical depth and number of clusters in the sample both with and without the luminosity cut.

\begin{table}[H]
  \begin{tabular}{ | l | cc | cc | }
    \hline
          & \multicolumn{2}{|c|}{With $L_X$ cut}            & \multicolumn{2}{|c|}{No $L_X$ cut} \\
    \hline
    $z_{max}$ & Number of Clusters & $\tau_{avg} \times 10^{3}$ & Number of Clusters & $\tau_{avg} \times 10^{3}$ \\
    \hline
0.16   & 131 & 6.3  & 573 & 3.7 \\
0.20   & 205 & 6.6  & 649 & 4.1 \\
0.25   & 269 & 7.0  & 713 & 4.5 \\
All z  & 327 & 7.0  & 771 & 4.9 \\
    \hline
  \end{tabular}
  \caption{Mean optical depth of our cluster sample.}
  \label{tbl:avgtau}
\end{table}

Although the average optical depth is higher with the luminosity cut, the noise (which is roughly proportional to $1/\sqrt{N_{\rm clusters}}$) is also higher and so we expect a similar sensitivity both with and without the luminosity cut. However, when we calculate the cluster dipole in maps filtered with our Wiener filter with our cluster sample we find much lower significance as shown in table~\ref{tbl:ourres}.

\begin{table}[H]
  \begin{tabular}{ | lll | cc | cc | cc | cc | }
    \hline
             &           &           & \multicolumn{8}{|c|}{Temperature [$\mu$ K]} \\
    \hline
    Cluster sample & $z_{min}$ & $z_{max}$ & m     & $\sigma_m$ & x    & $\sigma_x$ & y     & $\sigma_y$ & z     & $\sigma_z$  \\
    \hline
Ours & 0.00 & 0.16 & -1.58 & 0.45   &  -0.21 & 0.90   &  -1.75 & 1.09  &   -0.35 & 0.58 \\
Ours & 0.00 & 0.20 & -1.01 & 0.40   &  -0.08 & 0.68   &  -1.05 & 0.76  &   -0.51 & 0.48 \\
Ours & 0.00 & 0.25 & -1.15 & 0.34   &  -0.84 & 0.57   &  -0.98 & 0.68  &   -0.62 & 0.49 \\
    \hline
  \end{tabular}
  \caption{Result from our filter pipeline with the luminosity cut.}
  \label{tbl:ourres}
\end{table}

We conclude from these results that we can accurately reproduce the KAKE pipeline and do not detect a bulk flow with greater than $3\sigma$ significance in the WMAP maps. The significance of the results is lower when we use our Wiener filter and the full cluster sample than when using the KAKE filter and the cluster sample that has low luminosity clusters removed. By comparing with table~\ref{tbl:main_KY} we see that the results with the smaller cluster sample are more significant than when those with the full cluster sample. However, since we would expect any bulk flow to be of similar significance with either of the cluster samples we conclude that there is no significant detection of a bulk flow.

\end{document}